\documentclass[final,5p,times,twocolumn]{elsarticle}

\usepackage{amssymb}
\usepackage{amsmath}
\usepackage{lineno}
\usepackage{hyperref}
\usepackage{graphicx}	% Including figure files
\usepackage{ulem}
\usepackage{color}     % To cancel before submitting
\usepackage[dvipsnames]{xcolor}
\usepackage{svg}

\usepackage{natbib} 

\usepackage{listings}
\definecolor{clr-background}{RGB}{255,255,255}
\definecolor{clr-text}{RGB}{0,0,0}
\definecolor{clr-string}{RGB}{163,21,21}
\definecolor{clr-namespace}{RGB}{0,0,0}
\definecolor{clr-preprocessor}{RGB}{128,128,128}
\definecolor{clr-keyword}{RGB}{0,0,255}
\definecolor{clr-type}{RGB}{43,145,175}
\definecolor{clr-variable}{RGB}{0,0,0}
\definecolor{clr-constant}{RGB}{111,0,138} % macro color
\definecolor{clr-comment}{RGB}{0,128,0}

\definecolor{bg}{rgb}{1.0,1.0,1.0}              % White background
\definecolor{text}{rgb}{0.1,0.1,0.1}            % Main text
\definecolor{keywordcolor}{rgb}{0.0,0.0,0.6}    % Keywords (blue)
\definecolor{stringcolor}{rgb}{0.63,0.12,0.94}  % Strings (purple)
\definecolor{commentcolor}{rgb}{0.0,0.5,0.0}    % Comments (green)
\definecolor{numbercolor}{rgb}{0.5,0.0,0.0}     % Numbers (dark red)
\definecolor{rulecolor}{rgb}{0.8,0.8,0.8}       % Border gray

\lstdefinestyle{vscodewhite}{
    backgroundcolor=\color{bg},
    basicstyle=\ttfamily\scriptsize\color{text},
    keywordstyle=\color{keywordcolor}\bfseries,
    commentstyle=\color{commentcolor}\itshape,
    stringstyle=\color{stringcolor},
    numberstyle=\color{numbercolor},
    numbers=left,
    numbersep=10pt,
    frame=single,
    rulecolor=\color{rulecolor},
    breaklines=true,
    breakatwhitespace=true,
    showstringspaces=false,
    tabsize=4,
    captionpos=b,
    language=C++,
    morekeywords={constexpr,pragma,acc, update, host, device,paralle,loop,exit,data,copyout,parallel,enter,copyin,declare,create,Coor,Nebins,uint64_t,T}
}
\usepackage{comment}

\DeclareRobustCommand{\VAN}[3]{#2}
\let\VANthebibliography\thebibliography
\def\thebibliography{\DeclareRobustCommand{\VAN}[3]{##3}\VANthebibliography}

\usepackage{booktabs}
\journal{Astronomy and Computing}

\begin{document}
%\backgroundsetup{contents={\sffamily PREPRINT}}
%\linenumbers
\begin{frontmatter}



\title{The PLUTO Code on GPUs: Offloading Lagrangian Particle Methods}


\author[unito,inaf,icsc]{Alessio Suriano}
\author[unito]{Stefano Truzzi}
\author[inaf]{Agnese Costa}
\author[unito]{Marco Rossazza}
\author[cineca]{Nitin Shukla}
\author[unito]{Andrea Mignone}
\author[unito]{Vittoria Berta}
\author[inaf]{Claudio Zanni}

\affiliation[unito]{organization={UNITO - Università degli Studi di Torino,  Dipartimento di Fisica},%Department and Organization
            addressline={via Giuria, 1}, 
            city={Torino},
            postcode={10100},
            country={Italy}}
\affiliation[inaf]{organization={INAF - Istituto Nazionale di Astrofisica, Osservatorio Astrofisico di Torino},%Department and Organization
            addressline={Strada Osservatorio, 20}, 
            city={Pino Torinese},
            postcode={10025},
            country={Italy}}
\affiliation[icsc]{organization={ICSC - Italian Research Center on High Performance Computing, Big Data and Quantum Computing},%Department and Organization
            addressline={via Magnanelli, 2}, 
            city={Casalecchio di Reno},
            postcode={40033},
            country={Italy}}
\affiliation[cineca]{organization={CINECA},%Department and Organization
            city={Casalecchio di Reno},
            postcode={40033},
            country={Italy}}
%% Abstract
\begin{abstract}
%We present preliminary performance results of gPLUTO, the new GPU-optimized implementation of the PLUTO code for computational plasma astrophysics. Like its predecessor, gPLUTO employs a finite-volume formu-lation to numerically solve the equations of magnetohydrodynamics (MHD)in multiple spatial dimensions. Still, this new implementation is a completerewrite in C++ and leverages the OpenACC programming model to achieveacceleration on NVIDIA GPUs. While a more comprehensive description of the code and its several other modules will be presented in a companion paper, here we focus on some preparatory results that demonstrate the code potential and performance on pre exa-scale parallel architectures.

% One of the most compelling results of the modern observational astrophysics is the intense non-thermal radiation that characterizes many astrophysical sources.
% Various models have been proposed to explain such high-energy radiation, nevertheless the non-trivial behaviour of the astrophysical plasma, the interplay of phenomena acting on very different spatial scales, and the intrinsic observational challenges, still prevent us from pin-pointing a comprehensive acceleration scenario able to fully reproduce the observations. 
%In this context numerical experiments represent the only possibility to challenge our models by simulating the evolution of such extreme environments. 
%High-energy astrophysical sources present compelling non-thermal high-energy emission whose production mechanisms are still far from being deeply understood.
The Lagrangian Particles (LP) module of the \texttt{PLUTO} code offers a powerful simulation tool to predict the non-thermal emission produced by shock accelerated particles in large-scale relativistic magnetized astrophysics flows.
The LPs represent ensembles of relativistic particles with a given energy distribution which is updated by solving the relativistic cosmic ray transport equation.
The approach consistently includes the effects of adiabatic expansion, synchrotron and inverse Compton emission. 
%shock-driven particle acceleration in relativistic magnetized astrophysical flows.
The large scale nature of such systems creates boundless computational demand which can only be satisfied by targeting modern computing hardware such as Graphic Processing Units (GPUs).
In this work we presents the GPU-compatible C++ re-design of the LP module, that, by means of the programming model OpenACC and the Message Passing Interface library, is capable of targeting both single commercial GPUs as well as multi-node (pre-)exascale computing facilities.
The code has been benchmarked up to 28672 parallel CPUs cores and 1024 parallel GPUs demonstrating  $\sim(80-90)\%$ weak scaling parallel efficiency and good strong scaling capabilities.
Our results demonstrated a speedup of $6$ times when solving that same benchmark test with 128 full GPU nodes (4GPUs per node) against the same amount of full high-end CPU nodes (112 cores per node).
Furthermore, we conducted a code verification by comparing its prediction to corresponding analytical solutions for two test cases.
We note that this work is part of broader project that aims at developing \texttt{gPLUTO}, the novel and revised GPU-ready implementation of its legacy.

% This comes with the intrinsic cost of solving the relevant equations with a sufficient temporal and spatial resolution   the physical process responsible for large-scale systems ... brings.
% The latest hardware developments in Graphic Processing Units (GPU) design achieved unprecedented computational power...
% The Lagrangian Particle module of \texttt{PLUTO}, which aims at predicting the non-thermal emission of shock-accelerated particles in large-scale astrophysical systems, must be prepared to explore such modern architectures enhancing its value for high-energy astrophysics modelling.
% The goal is to bring to the community a high-performance scalable and portable code that allows for large  high resolution, simulation.

% We present the complete redesign of the Lagrangian Particles module of the PLUTO code to target the modern (pre-)exhale GPU-capable High Performance Computing clusters.

\end{abstract}


% \begin{highlights}
% \item blah
% \end{highlights}

%% Keywords

\end{frontmatter}

\section{Introduction}

High-energy radiation in the Universe originates from several classes of astrophysical systems, including Active Galactic Nuclei, Gamma-ray Bursts, Supernova Remnants, Pulsar Wind Nebulae, and X-ray Binaries. All of these sources share distinctive non-thermal radiation signatures that can be detected across a wide range of energies, from radio waves to gamma-rays \cite{blandford_konigl1979,kennel_coroniti1984b}.%,mirabel_rodriguez1999,kumar_zhang2015}. 

Such multi-wavelength radiation is produced by non-thermal particles (NTPs) whose acceleration mechanisms are still debated.
The two most promising acceleration candidates are relativistic magnetohydrodynamic (RMHD) shocks \citep{krymskii1977,axford1977} and magnetic reconnection, \citep{kulsrud1998} given their capability of producing power-law distributions of particles consistent with observations and their ubiquity in astrophysical plasmas. %\cite{dalpino2010,drake2013,caprioli_spitkovsky2014a}.
Other possible acceleration scenarios include Fermi II and shear \cite{fermi1949,rieger_duffy2004}.
Current observations are only able to indirectly probe these phenomena by detecting their by-products in the form of electromagnetic radiation that reaches our instruments after being absorbed and re-emitted along the path.
%, observations are often difficult to interpret, as they are limited to the radiation emitted and absorbed by these systems, leaving many plasma processes hidden from direct measurement.
Moreover, the study of astrophysical plasmas is made complex by the wide range of spatial and temporal scales on which these fundamental processes %governing the radiation we detect 
take place. 
These scales range from microscopic ones, such as the electron / ion skin depth or the Larmor radius, to macroscopic dynamical scales  %that determine the conditions at particle acceleration-emission sites, as well as beaming effects \cite{uzdensky2006}, 
that reach up to megaparsec distances in the case of relativistic jets.
Thus, when coming to high-energy astrophysical sources, there is still no general consensus regarding structure formation processes, intrinsic morphology, and the matter-energy balance, as well as the emission channels and acceleration mechanisms underlying their non-thermal signatures.

%In this context, simulations play a key role acting as numerical experiments: they allow us to model the system under controlled conditions and to probe regimes inaccessible to laboratory experiments.
%By directly comparing synthetic results with observations, they establish a crucial connection between theory and observational data.
%However, the intrinsically multi-scale nature of astrophysical systems remains a significant challenge.
In this picture, numerical simulations play a key role allowing for system modelling under controlled conditions and offering the possibility of probing regimes inaccessible to laboratory experiments.
Such numerical experiments make possible to directly relate the simulated ambient conditions to physical observables.
Furthermore, the comparison of the synthetic data with astrophysical measurement offers an excellent test ground to challenge the predictions of the theoretical model.
%Furthermore by directly comparing synthetic results with observations, they establish a crucial connection between theory and observational data.
%Furthermore they establish a direct connection between  to  by directly comparing synthetic results with observations, they establish a crucial connection between theory and observational data.

However, the intrinsically multi-scale nature of astrophysical systems remains a significant challenge.
Fluid simulations, solving the (R)MHD equations \cite{davidson2001}, cannot capture or resolve the plasma kinetic scales responsible for particle acceleration.
On the other hand, Particle-in-Cell (PIC) simulations provide a more consistent description of particle dynamics by resolving the electron Larmor radius \cite{birdsall1991}, but they cannot account for large scale plasma evolution, because of the huge numerical costs required to simulate a statistically significant number of particles \cite{ji2022}. 

To bridge the gap between micro- and macro-scale descriptions, hybrid frameworks %can be developed
that operate across multiple scales have been developed.
There approaches combine the scalability of fluid models with the capacity to capture specific physical processes of interest.
In this respect, a widely used method for capturing kinetic effects is the kinetic ions-fluid electrons model, in which the former are treated as particles using PIC techniques, while the latter are modelled as a charge-neutralizing fluid \cite{lipatov2002}. %winske1985,guo_giacalone2013,caprioli_pop_spitkovsky2014,haggerty_caprioli2019}.%,(e.g., Byers et al. 1978; Hewett & Nielson 1978,e.g., Kunz et al. 2014; Bott et al. 2021; Jain et al. 2022).
Another approach is the MHD-PIC model, that distinguishes between the thermal plasma, described by (R)MHD equations, and NTPs, which are advanced using PIC methods \cite{mignone2018}.
This strategy is particularly well-suited for modelling cosmic-ray acceleration and transport, as well as their feedback on astrophysical flows.
Similarly, other hybrid approaches embed localized PIC domains within global fluid simulations to capture specific kinetic effects, such as magnetic reconnection or collisionless shocks, within large-scale environments \cite{daldorff2014,makwana_etal2017}.

%\ASc{I suggest a modification, Agnese correct me if i am wrong}
%\AS{Even though these approaches have the advantage of reducing the computational load by deferring system evolution to the solution of fluid-like equations, they must distinguish individual particles motion.
%When modelling extended astrophysical objects, the typical spatial and temporal scales of the bulk are much higher than the gyroradius and gyration time of the NTPs responsible for the high-energy emission.}
At even larger spatial and temporal scales, where the particle gyroradius is much smaller than the grid cell size, it is convenient to combine the Eulerian (R)MHD description of the thermal fluid with a sub-grid Lagrangian prescription for NTPs behaviour.
%where the LPs are sub-grid clouds of NTPs
This approach consist in advecting the NTPs on the Eulerian grid evolving their energy-space distribution based on the solution of Fokker-Planck transport given the in situ fluid conditions.
In this way, it is possible to account for a variety of physical processes, including adiabatic losses, non-thermal cooling, diffusive shock acceleration, and related effects.
%Here, unlike the MHD-PIC model, the non-thermal populations do not exert a beck-reaction on the system .
%In this approach, the spectral distribution of Lagrangian particles (LPs) is governed by a Fokker-Planck equation that accounts for a variety of physical processes, including adiabatic losses, non-thermal cooling, diffusive shock acceleration, and related effects.
%At the numerical level, this is achieved by locally sampling the fluid conditions and employing semi-analytical methods to evolve the particle distribution function in time.
%\cite{kardashev1962,gomez_etal1995,murgia_etal1999,aloy_etal2000,shibata_etal2003,komissarov_lyubarsky2004,bottcher_dermer2010,porth_etal2011,hardcastle_krause2014}. 

In classical fluid dynamics, the Lagrangian–Eulerian framework has been used to model diffusive shock acceleration alongside adiabatic and synchrotron cooling \cite{jones_etal1999,micono_etal1999}.
This technique was later extended to relativistic HD and MHD codes \cite{mimica_etal2009}
%,delacita_etal2016,fromm_etal2016,vaidya2018,ogrodnik_etal2021,larissa_etal2025}
, and it was incorporated into the Lagrangian Particles (LP) module of the \texttt{PLUTO} code for astrophysics \cite{mignone2007,vaidya2018} through a novel approach that predicts the non-thermal power spectrum of NTPs accelerated in (R)MHD shock.
This implementation is not only able to incorporate the dependence of the spectral index on the shock strength and magnetic field orientation, but is also capable to take into account adiabatic energy losses, synchrotron emission, polarization, and inverse Compton scattering.
More recent developments \cite{mukherjee2021} made possible to keep track of the particles acceleration history, without neglecting the effects due to multiple accelerations \cite{micono_etal1999}.
The module, and its subsequent extensions, has since been employed in various cases, particularly for modelling relativistic jets
\cite{giri_etal2022,kundu_etal2022,meenakshi_etal2023,sciaccaluga_etal2025}.

While hybrid Lagrangian–(R)MHD simulations build on progress in fluid modelling, the simultaneous treatment of the fluid and particle components greatly increases computational requirements. 
Recent advances in high-performance computing, especially the widespread use of GPU-accelerated codes in astrophysical plasma studies \cite{shukla2025}, now make such hybrid approaches increasingly feasible.

This work presents the redesign of the LP module to target GPU offloading in the context of \texttt{gPLUTO}, the GPU-ready implementation of its legacy \texttt{PLUTO} presented in a companion article \cite{rossazza2025}.
The paper is structured as follows: in \S \ref{sec:numerical_methods} we provide a description of the numerical method, while in \S \ref{sec:gpu_design} technical details on the GPU implementation are presented.
\S \ref{sec:benchandtest} is dedicated to the numerical benchmarks and parallel performance, while \S \ref{sec:conclusion} summarizes our results.
%%%%%%%%%%%%%%%%%%%%%%%%%%%%%%%%%%%%%%%%%%%%%%%%%%%%%%%%%%%%%%%%%%%%%%
\section{Numerical methods}
\label{sec:numerical_methods}
%%%%%%%%%%%%%%%%%%%%%%%%%%%%%%%%%%%%%%%%%%%%%%%%%%%%%%%%%%%%%%%%%%%%%%

A Lagrangian Particle (LP) represents a cloud of NTPs (ions or electrons) that are very close in physical space so that they can be approximated to a Dirac delta in the Eulerian grid at position $\textbf{x}_{\rm LP}$.
% The macroparticle approach samples the spatial and temporal evolution of the distribution function
As the LPs are advected according to %on the PLUTO grid according to
\begin{equation}\label{eq:push}
  \frac{d \textbf{x}_{{\rm LP}}}{dt} = \textbf{v}(\textbf{x}_{\rm LP})\, ,
\end{equation}
% 
%where $\textbf{x}_{\rm LP}$ is the LP position, $\textbf{v}$ is the fluid velocity interpolated at $\textbf{x}_{\rm LP}$.% and the subscript \quotes{p} labels the individual MP.
they sample fluid quantities such as bulk velocity $\textbf{v}$, magnetic field, density and pressure.
Note that these fluid quantities are evolved concurrently by solving the (R)MHD conservation equations. 
The NTPs composing the LPs have a non trivial phase-space distribution function $f_0(p, \tau)$ which assumed to be isotropic in momentum space and therefore depends only on the magnitude of the momentum $p$. % and on the LP proper time $\tau$.
%, assumed to be isotropic in momentum, thus only depending on the magnitude of the momentum $p$.
Here $\tau$ is the proper time of the LP and is related the laboratory time by $d \tau = dt / \gamma_{\rm g}$ and $\gamma_{\rm g}$ denotes the bulk (fluid) Lorentz factor.
Note that since we treat relativistic particles we will refer to the energy space (rather than momentum space) distribution given that $\epsilon\approx pc$ where $c$ is the light speed.
The relativistic cosmic-rays transport equation \cite{Webb1989} prescribes the evolution of the energy-space distribution function away from shocks:
% the distribution function
% \begin{equation}
%     \label{eq:f_distro}
%     f_0 (x^\mu, p) \,,
% \end{equation}
% %
% where $x^\mu$ is the position four-vector while $p$ is the magnitude of the momentum. 
% Note that $f_0$ depends only on the magnitude of the momentum as we assume isotropy in momentum space.
% Eq. (\ref{eq:f_distro}) satisfies the relativistic cosmic-rays transport equation \cite{Webb1989} which, neglecting spatial diffusion, shear and turbulent accelerations, reads

%
\begin{equation}
    \label{eq:cr_N}
    \frac{d \mathcal{N}}{d \tau} + \frac{\partial}{\partial \epsilon} \left[ \left( - \frac{\epsilon}{3} \nabla_\mu u^\mu + \frac{d\epsilon_{\rm l}}{d\tau} \right) \mathcal{N} \right] = - \mathcal{N} \nabla_\mu u^\mu\,,
\end{equation}
where $u^\mu$ is the bulk four-velocity of the fluid and
%$\mathcal{N}\equiv \mathcal{N}(\epsilon,\tau)$ 
%
\begin{equation}
    \label{eq:n_epsilon_def}
    \mathcal{N}\equiv\mathcal{N}(\epsilon, \tau)=\int d \Omega p^{2} f_{0} \approx 4 \pi p^{2} f_{0} \,
\end{equation}
is the number density of actual particles particles with energy between $\epsilon$ and $\epsilon + d\epsilon$.
%particles depends on to the momentum $p$, assumed isotropic, and on the distribution function $f_0$.
%In the relativistic limit the energy of a particle can be expressed as $\epsilon \approx pc$, therefore $\mathcal{N}(\epsilon, \tau) d \epsilon = \mathcal{N}(p, \tau) dp$. 
The two terms in the square bracket in Eq. (\ref{eq:cr_N}) account, respectively, for adiabatic expansion and radiative losses due to synchrotron and inverse Compton emissions \citep[see][]{vaidya2018,mukherjee2021}:
\begin{equation}
    \label{eq:rad_losses}
    \frac{d\epsilon_{\rm l}}{d\tau} = - c_{\rm r} \epsilon^2 \qquad 
    {\rm where}\qquad
    c_{\rm r} = \frac{4}{3} \frac{\sigma_{\rm T} \beta_{\rm e}^2}{m_{\rm e}^2 c^3} \left[ U_{B} + U_{\rm rad}  \right] \,,
\end{equation}
while $U_{B}$ and $U_{\rm rad}$ are the energy densities of the local magnetic and radiation fields, $\sigma_{\rm T}$ is the Thompson cross section, $m_{\rm e}$ is the electron mass and $\beta_{\rm e}\approx 1$ is the velocity of relativistic electrons.

By defining the number density fraction of non-thermal particles with respect to the fluid particles number density $n$
\begin{equation}
    \label{eq:chi}
    \chi (\epsilon)=\frac{\mathcal{N}(\epsilon)}{n}
\end{equation}
% \begin{equation}
%     \chi (\epsilon) = \mathcal{N}(\epsilon) /n \quad ,
% \end{equation}
% {\MAGENTA COMMENTO: forse si puo' mettere come testo invece che come equazione}
% %
and using the continuity equation $\nabla_\mu(nu^\mu)=0$, we can rewrite Eq. (\ref{eq:cr_N}) as
\begin{equation}
    \label{eq:cr_chi}
    \frac{d \chi}{d \tau} + \frac{\partial}{\partial \epsilon} \left[ \left( - \frac{\epsilon}{3} \nabla_\mu u^\mu + \frac{{d \epsilon}_{\rm l}}{d \tau} \right) \chi \right] = 0 \,.
\end{equation}

When a LP crosses a shock, the out-coming power spectrum is updated in relation to the in-coming one as
\begin{equation}
\begin{aligned}
\label{eq:convolution_spectra}
    \chi_{\rm out} (\epsilon) &= 
    \mathcal{C} \int_{\epsilon_{\min}}^\epsilon \chi_{\rm in}(\epsilon') \chi_{\rm sh} (\epsilon,\epsilon') \frac{\mathrm{d} \epsilon'}{\epsilon'} \\
    &= \mathcal{C} \int_{\epsilon_{\min}}^\epsilon \chi_{\rm in} (\epsilon') \left( \frac{\epsilon}{\epsilon'} \right)^{-q+2} \frac{\mathrm{d}\epsilon'}{\epsilon'} ,
\end{aligned}
\end{equation}
where $\mathcal{C}$ is a normalization that depends on the shock energy (see \cite{vaidya2018,mukherjee2021} for further details) and the power--law index 
\begin{equation}
\label{eq:slope}
    q(r)=\frac{3r}{r-1} + q_{\rm R}(r,\beta_{\rm u,d})
\end{equation}
is a function of the compression ratio $r$ only for non-relativistic shocks.
In the case of a relativistic fluid, the correction $q_{\rm R}$, which depends on the plasma velocities upstream and downstream of the shock $\beta_{\rm u, d}$, is applied \cite{drury1983,keshet2005,takamoto2015}.
The numerical solution of Equations \ref{eq:chi} and \ref{eq:convolution_spectra} is found by discretizing the spectrum over $N_{\rm b}$ bins (see Section 2.3.2 of \cite{vaidya2018}).

\section{Exploiting Graphic Processing Units}
\label{sec:gpu_design}
%%%%%%%%%%%%%%%%%%%%%%%%%%%%%%%%%%%%%%%%%%%%%%%%%%%%%%%%%%%%%%%%%%%%%%%%%%

%\subsection{Why GPUs}
The \texttt{gPLUTO} code has been developed and it is actively maintained to provide a high-level yet user-friendly framework, enabling both entry-level and experienced researchers to use it effectively with minimal setup.
Its portability among a variety of computing architectures together with its scalability capabilities have been key features in increasing its popularity within the computational astrophysics community.
Maintaining its leading role requires adapting the code implementation to modern (pre-) exascale computing infrastructures, which provide high levels of parallelism through Graphics Processing Units (GPUs).

GPUs are specialized electronic hardware developed for efficient digital image processing that, since their initial conception, have evolved to become powerful general-purpose computing engines; a transition driven by their clear and massive parallel processing potential.
This has made their usage fundamental for achieving high performance in scientific applications.
A GPU's computing power is derived from its architecture, which consists of hundreds to thousands of simplified, highly parallel cores.
%These cores are engineered for throughput, efficiently managing simultaneous execution of numerous threads, enabling to process vast datasets in parallel.
Each of these cores are engineered to work synchronously by executing the same instruction on different data, making them particularly suitable for the mesh-based implementation of the (R)MHD equation solvers implemented in PLUTO as well as for the LPs.

{In the context of a NVIDIA GPU architecture,} the computational work unit executed on the GPU (or device) is defined by the kernel, a code block compiled and launched from the CPU (host).
The device organizes the execution of a kernel into a structured hierarchy:
% \begin{itemize}
%     \item Grid and Blocks: The kernel is launched across a grid, which is a collection of independent thread blocks.
%     \item Threads: Each block is composed of a fixed number of threads, which are the smallest units of execution carrying out a single stream of program instructions. Threads within the same block execute concurrently on a single Streaming Multiprocessor (SM), which is the foundational building block of the NVIDIA GPU architecture.
%     \item Shared Resources: Threads within a block can communicate and share data efficiently via a pool of high-speed Shared Memory, and they can synchronize their execution.
% \end{itemize}

\begin{enumerate}
    \item Thread - The smallest units of execution carrying out a single stream of program instructions.
    \item Warp - Group of threads that are executed in parallel. It is the fundamental scheduling unit as the threads of a warp execute the same instruction simultaneously. Note that minimizing loop divergences is crucial since when the threads in a warp split from one another to execute different instructions, the performance drops precipitously.
    \item Block - The smallest unit of threads coordination exposed to programmers. When a kernel is launched, one or more thread blocks are produced, each of which contains one or more warps. They can be arbitrarily sized, but they are typically multiples of the warp size.
    \item Grid - The kernel is launched on a set of multiple blocks that compose a grid. The grid blocks can be executed with different degrees of concurrency depending on the device architecture.
\end{enumerate}
The blocks of a given kernel are scheduled and executed on Streaming Multiprocessor, the physical self-contained processing unit that composes a GPU.
They are equipped with individual register and L1 memory cache and are able to work independently form the other SMs of the device.
%All threads execute the same program, but they operate on different data located at unique memory addresses. To maximize efficiency and memory access performance, the SM organizes thread execution into warps, where a warp is a fundamental scheduling unit consisting of 32 threads. Grouping threads into warps allows the hardware to optimize data access: threads within a warp are encouraged to access consecutive memory addresses (coalesced memory access), significantly reducing memory latency and improving overall computational throughput.

\subsection{Programming model}

{
To offload \texttt{gPLUTO} to GPUs without sacrificing simplicity and portability, we adopted OpenACC \cite{openacc}, a high-level programming model introduced in 2011 by Cray, CAPS, NVIDIA, and PGI. While current compiler support effectively restricts our implementation to NVIDIA accelerators, this choice provides a solid foundation for future extensions, including the exploration of alternative models such as OpenMP for non-NVIDIA platforms.
}

%OpenACC allows the code to target { NVIDIA} parallel accelerators through compiler directives.

OpenACC makes use of compiler directives and clauses to instruct device actions:
\footnotesize
\begin{lstlisting}[
    label={lst:pragma},
    style=vscodewhite,
    ]
#pragma acc <directive> <cl_1, cl_2, ...>
\end{lstlisting}
\normalsize
To run the following copy loop on a GPU it is sufficient to trigger host-device copies of the relevant data (lines 1-2, 9-10) and instruct the kernel launch by decorating the loop with the compute directive (line 4).
\begin{lstlisting}[
    label={lst:kernel},
    style=vscodewhite,
    ]
#pragma acc enter data copyin(b[:N])
#pragma acc enter data declare create(a[:N])

#pragma acc parallel loop
for (int i = 0; i < N; i++){
  a[i] = b[i]
}

#pragma acc exit data delete(b[:N])
#pragma acc exit data copyout(a[:N])
\end{lstlisting}
Note that OpenACC directives are ignored when compiled on CPU-based architectures, ensuring complete cross-platform compatibility.
% could have a non unique The These two processors have separate memories, and that the operation special care has to be posed in allocating the data in the whether the data are store
% We remark that the activities of the accelerator are coordinated by the CPU, meaning that  
% In accelerating an application particular care has to be posed on data residency. The program  Note that for a given application, its
% Note that the data used inside a parallel region have to be previously copied into the device (GPU) memory and that the results will reside 
% Pragma directives are also used to manage the transfer of data between the host (CPU) and the device (GPU) memory.
%The latter is of key importance ensure data coherency end result correctness as the
%When offloading the application, the CPU coordinates the execution, taking care of initialization, I/O operations and kernel launching, whereas the entire problem evolution is performed by the GPU.
%When running on accelerated systems (i.e. equipped with GPUs), the main bottleneck is host-device data movement.
%This comes from the fact that production GPUs are extremely efficient in performing double--precision floating--point operations to the point that any data movement or memory operations (i.e. allocation/deallocation) 
%\subsection{Kernels optimization - coalescent memory access}
% \subsection{Porting cycle}
% \ASc{Da fare forse, @Alessio, @Marco}
\subsection{Performance optimization strategies}
When designing code and data structures we consider the following key best practices.
\paragraph{Reduce host-device data transfer}
The activity of the GPU accelerators must always be initialized, coordinated, and finalized by the host, meaning that the program operates in a hybrid fashion across the two platforms.
Furthermore, since the host and device have separate memory spaces, data residency must be explicitly managed by ensuring that input data is present on the device before kernel execution, and that the results are copied back to the host afterward.
To reduce the latency introduced by host-device communication, we ensure device data locality by executing all the computations in accelerated regions, deferring host data updates until I/O operations are required.
{
We remark that hardware vendors are currently investing significant effort in overcoming such sharp division between host and device memory spaces.
The emergence of coherent CPU–GPU memory architectures has the potential to reduce the complexity and inefficiencies associated with explicitly managing data locality.
Nevertheless, careful data-locality management remains important to ensure full compatibility with non-coherent architectures, which represent the most common layout for the currently available EuroHPC clusters.
}
\paragraph{Minimize memory allocation/deallocation}
Allocating or deallocating large chunks of memory must be limited to the greatest extent as it is associated to significant overheads.
Notably, within the LP module, the number of particles per processor is not constant during the simulation as 
%The particles at run-time move either at physical or at computational boundaries and are have to be deleted when they are advected over the physical domain or the inter-processor interfaces.
%Notably, within the framework of the LP module, the particle count per processor is not constant over the course of a simulation.
they can be injected run-time, they can be advected outside the physical boundaries or they can cross inter-processors interfaces.
The LP module accommodates such dynamical behaviour concurrently ensuring a minimal impact on the application performance by implementing a chunked memory allocation.
In particular, given a number of particles to be stored \texttt{N\_} and a length of the chunk \texttt{L\_ = $2^{e}$}, we allocate a number of chunks
\begin{lstlisting}[
    caption=Number of chunks evaluation,
    label={lst:chunksize},
    style=vscodewhite
    ]
Nc_ = (N_ >> e) + (N_ % L_ == 0 ? 0 : 1 );
\end{lstlisting}
We refer the reader to \ref{sec:containers} for further details on the implementation.

\paragraph{Ensure coalesced access to memory}
\label{sec:mem-manag}
For devices with Compute Capability 6.0, when the threads of a warp access contiguous memory locations, the memory transaction is executed at the cost of a single 32-byte operation, resulting in optimal bandwidth \cite{cudaBestPractise}.

In general this is achieved by guaranteeing an ordered and linear indexing, where the index $i$ of a loop accesses the $i$-th item of an array.
In the case of the present discussion, where each particle is an entity characterized by a set of variables (e.g. position, velocity, energy spectrum, etc.), providing coalesced memory means to allocate a structure of arrays (rather than an array of structures).
This reads that, for a given number of particles $N_{\rm p}$ characterized by a set of $n$ parameters, we allocate $n$ arrays with size $N_{\rm p}$ where the element with index $i=0,..,N_{\rm p}-1$ of each array is a parameter of the $i$-th particle.

Note that, since particles are not stored in the structure in any particular order, element deletion has to be associated to a compaction strategy to ensure data continuity in memory.
To this purpose we developed the following algorithm:
%In particular we overwrite the values corresponding to the $n_{\rm del}$ dying particles with the ones of the $n_{\rm del}$ trailing ones.
\begin{enumerate}
    \item Flag and count the number $N_{\rm del}$ of particles that have to be deleted (holes).
    \item Save the indices $i_{\rm del}$ of the flagged particles in a temporary array. This array will had been previously resized so that its size $N_{\rm tmp} \geq  N_{\rm p}$.
    \item If at least one particle index has been saved, collect the indices $i_{\rm f}$ of the fillers, i.e. the particles stored in position $i > N_{\rm p} - N_{\rm del}$ that do not have to be deleted.
    \item Copy each parameter of the particles with index $i_{\rm f}$ to the holes $i_{\rm del}\leq N_{\rm p}- N_{\rm del}$ collected at step 2.
    \item Shrink the structure so to satisfy condition of Listing \ref{lst:chunksize}.
\end{enumerate}
The algorithm uses two cycles with size $N_{\rm p}$, one with size $N_{\rm del}$ and one with size $N_{\rm f}$ which leads to an overall linear complexity $\mathcal{O}(N_{\rm p})$ and memory usage $\mathcal{O}(N_{\rm p})$.

\subsection{Multi-core communication}
\texttt{gPLUTO} addresses large-scale, high-resolution physical problems by evenly distributing the computational grid cells across parallel cores.
The synchronization of the fluid state in the cells forming the inter-core interfaces is based on the routines of the Message Passing Interface (MPI) library \cite{mpi}.
On top, LPs have to be exchanged when they are advected across the core interfaces.
The number of exchanged particles may vary run-time, requiring the adaptation of the communication buffer sizes as well as the particle containers capacity.
In this context, particle exchange is performed in three steps for the three parallel directions.
For every direction we can identify a left and a right processor and proceed in the following way:
\begin{enumerate}
    \item Collect holes and fills according to the procedure of Section \ref{sec:mem-manag} and count how many particles have to be sent to a given rank.
    \item Initialize \texttt{MPI\_Irecv()} for the incoming particles.
    \item Resize send/receive buffers to accommodate exchanging particles.
    \item Pack particles into the buffers.
    \item Start non-blocking \texttt{MPI\_Isend()}.
    \item Resize the particle container to accommodate for the new particles.
    \item Finalize send/receive calls.
    \item Copy particles from the buffers in to the holes left by outgoing particles and compact the structure.
\end{enumerate}
The use of non-blocking MPI calls allows to overlap left and right exchanges as well as the memory operation needed to resize the local particle container.
We plan to further extend the level of parallelism of the present algorithm  in future work by allowing simultaneous exchange of particles in all directions.
\subsubsection{Code sections overview}
We summarize here the routines the compose one simulation time-step together with their timing measured by profiling a single node run of the advection configuration described in Section \ref{sec:performances} on Leonardo Booster.
In this context and during the application porting we made used of the NVIDIA profiling tool \texttt{nsys}.
As reported in Table \ref{tab:codeTiming} the particles are advected in two Runge-Kutta (RK) sub-steps that take respectively $70$ms and $81$ms, whereas the most time-consuming routine is the one relative to the spectral update.
The time-step includes two accessory function: \texttt{Particles\_Boundary()}, which is responsible for the application of physical boundary conditions at the domain edges, and \texttt{Particles\_Exchange()}, that encapsulates the inter-core particle exchange algorithms.

\begin{table}[h!]
\centering
\begin{tabular}{l r}
\toprule
\textbf{Function}  & \textbf{Time (ms)} \\
\midrule
\texttt{Particles\_RK\#1()} & 70 \\
\texttt{Fluid\#1()}  & 32 \\
\texttt{Particles\_RK\#2()} & 81 \\
\texttt{Particles\_Boundary()} & 2 \\
\texttt{Particles\_Exchange()} & 18 \\
\texttt{Fluid\#2()}  & 32 \\
\texttt{Particles\_Spectra()}  & 190 \\ \midrule
\textbf{Total}  & 425 \\\midrule
\bottomrule
\end{tabular}
\caption{Functions that compose one \texttt{gPLUTO} time step with relative GPU execution timings in milliseconds. The timings are measured on the $100^{\rm th}$ time step of the shock configuration described in section \ref{sec:performances} that has been run on one full Leonardo Booster node.}
\label{tab:codeTiming}
\end{table}

% \begin{figure*}[h!]
%     \centering
%     \includegraphics[width=\textwidth]{figures/CodeProfiling3.png}
%     \caption{\AS{Describe the figure better}Nsys profiling showing the relevant sections composing a time step.}
%     \label{fig:codeprof}
% \end{figure*}

%Note that even though the focus of the present work is on Lagrangian Particles, the implementation here presented is versatile and applicable to any particle type present in the legacy code, PLUTO, and it is ready to be extended in future porting works. 
%%%%%%%%%%%%%%%%%%%%%%%%%%%%%%%%%%%%%%%%%%%%%%%%%%%%%%%%%%%%%%
\section{Numerical Benchmarks and Performance Assessment}
\label{sec:benchandtest}
%%%%%%%%%%%%%%%%%%%%%%%%%%%%%%%%%%%%%%%%%%%%%%%%%%%%%%%%%%%%%%

\subsection{Test Cases}
\label{sec:test}

%%%%%%%%%%%%%%%%%%%%%%%%%%%%%%%%%%%%%%%%%%

We present here two numerical benchmarks to validate the correctness of our numerical implementation.

%%%%%%%%%%%%%%%%%%%%%%%%%%%%%%%%%%%%%%%%%%%%%%
\subsubsection{Simple Advection}
%%%%%%%%%%%%%%%%%%%%%%%%%%%%%%%%%%%%%%%%%%%%%%

In our first test we assess the accuracy of the LP pusher by considering a 2D square and periodic Cartesian domain defined by $x,y\in[-1,1]$.
Density and pressure are set to unity throughout the domain while a velocity profile $\mathbf{v}=(v^\prime \cos{\theta}, v^\prime \sin{\theta})$ is prescribed.
Here $v^\prime=2 + \sin(\pi x^\prime)$ where $x^\prime = x\cos(\theta) - y\sin(\theta)$ is the abscissa of a Cartesian reference system rotated by an angle $\theta=\frac{\pi}{2}$ with respect to the PLUTO domain.
We freeze the fluid evolution to its initial status and we create 1 particle at $x=0$.

\begin{figure}[!ht]
    \centering
    \includegraphics[width=\columnwidth]{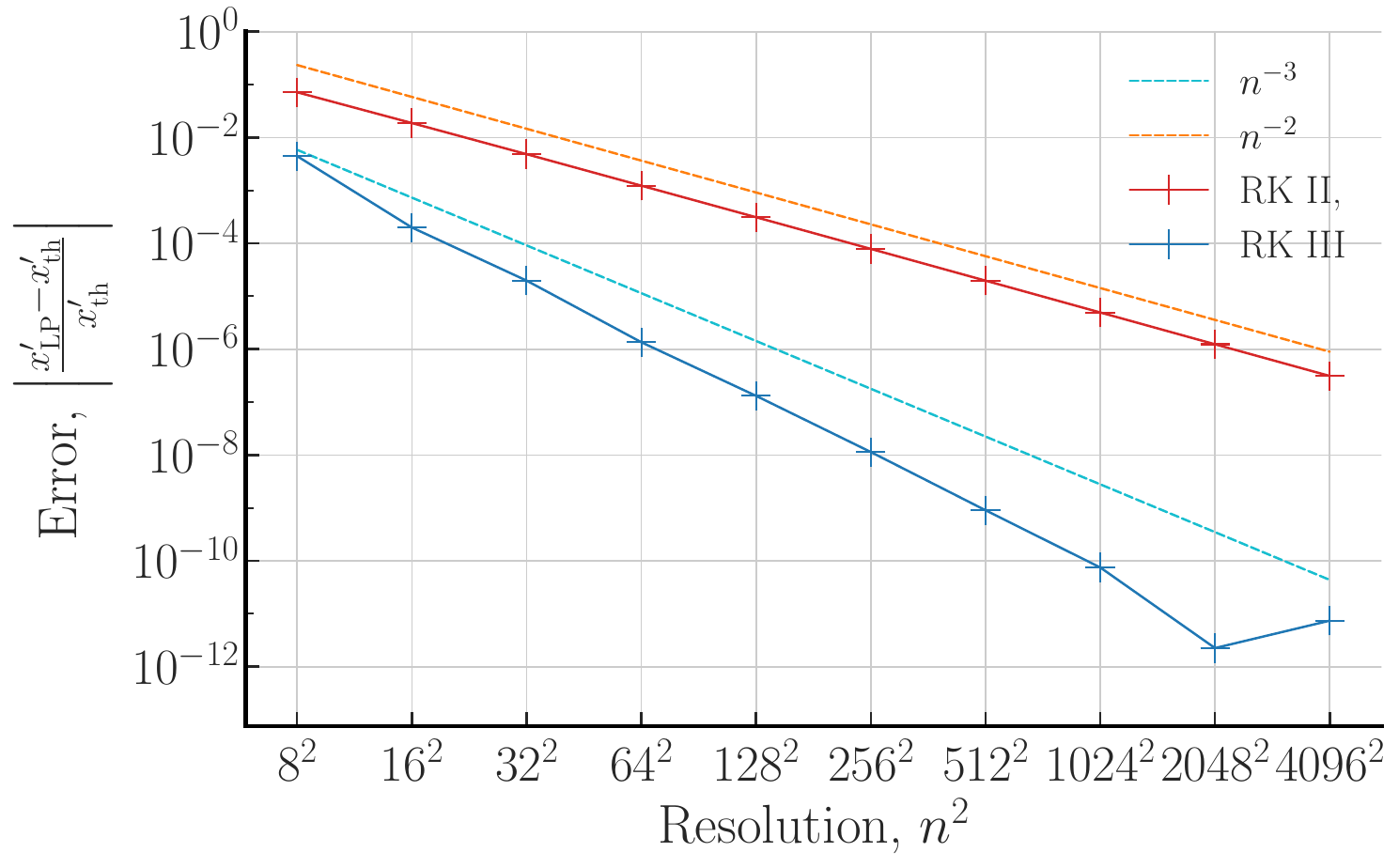}
    \caption{\footnotesize Errors for the particle advection problem at $t=10$, computed as the absolute { relative} difference between the LP position $x_{\rm LP}^\prime$ and the expected value $x_{\rm th}^\prime$, obtained from Eq. (\ref{eq:advection_exact_sol}).
    Red and blue solid lines correspond, respectively, to the $2^{\rm nd}$ and $3^{\rm rd}$ order Runge-Kutta algorithms, while dashed lines gives the expected accuracy.}
    \label{fig:advect}
\end{figure}

In Figure \ref{fig:advect} we plot the absolute { relative} difference between the position of the LP in the rotated system $x^\prime_{\rm LP}$ at $t=10$ and the theoretical position $x^\prime(t=10)\equiv x^\prime_{\rm th}$ as a function of the grid resolution for the second and third order Runge-Kutta time marching schemes respectively.
Here $x^\prime_{\rm th}$ is recovered analytically by solving
\begin{equation}
\begin{gathered}
\label{eq:tofx}
  t = \int \frac{dx^\prime}{2+\sin(\pi x^\prime)}\\
  =\frac{2\sqrt{3}}{3\pi}\arctan\Bigg(\frac{\sqrt{3}}{3}\Big(2\tan\Big(\frac{\pi x^\prime}{2}\Big) + 1\Big)\Bigg) + C\,,
\end{gathered}
\end{equation}
where we set $C=-\frac{1}{3\sqrt{3}}$ to impose the initial condition $t=0$ when $x^\prime=0$.
By inverting Equation \ref{eq:tofx} we obtain the expression
\begin{equation}\label{eq:advection_exact_sol}
  x^\prime(t) = \frac{2}{\pi}\arctan\Bigg[\frac{1}{2}\Bigg(\sqrt{3}\tan\Big(\frac{\pi(\sqrt{3} + 9t)}{6\sqrt{3}}\Big)-1\Bigg)\Bigg ].
\end{equation}

We observe that the accuracy improves with increasing resolution, and in both cases the observed trends are consistent with the expected accuracy scaling of $n^{-2}$ and $n^{-3}$ for the $2^{\rm nd}$- and $3^{\rm rd}$-order Runge–Kutta schemes, respectively.

\begin{comment}

In Figure \ref{fig:advect} we show the speed of one particle as a function of its displacement, { [defined as ?]} or the distance travelled along the $x$ axis since the beginning of the simulation.
The velocity of the LP follow a sinusoidal function with periodicity $\Pi=2$, amplitude $A=1$ and offset $B=2$, matching the fluid velocity field prescription.
{ [THIS IS A LAME STATEMENT... IT WORKS BETTER IF WE PROVIDE QUANTITATIVE EVIDENCE]}

\label{sec:advTest}
\begin{figure}[h!]
    \centering
    \includegraphics[width=\columnwidth]{vel1.0.pdf}
    \caption{Speed of one particle as a function of its displacement w.r.t its initial position. Note that the displacement can be greater than the domain size thanks to the periodic boundary conditions.}
    \label{fig:advect}
\end{figure}
\end{comment}

%%%%%%%%%%%%%%%%%%%%%%%%%%%%%%%%%%%%%%%%%%%%%%
\subsubsection{Stationary Planar Parallel Shock}
\label{sec:shockTest}
%%%%%%%%%%%%%%%%%%%%%%%%%%%%%%%%%%%%%%%%%%%%%%
In the $2^{\rm nd}$ benchmark we evaluate the spectral update algorithms in the presence of an MHD shock.
We solve the equations in the shock rest frame on a 2D square domain with $x,y\in[-1,1]$ using $128^2$ grid zones.
The shock is placed at $x=0$.
Downstream fluid quantitates ($x>0$) are related to the upstream density $\rho_{u}=50$, pressure $p_{u}=0.01$ and velocity $v_{x,u}=0.1$ according to the Rankine–Hugoniot conditions
\begin{equation}
\left\{\begin{array}{lcl}
    \rho_d  &=&\displaystyle \frac{\rho_u}{r}  \,,
      \\ \noalign{\medskip}
    p_d     &=&\displaystyle p_u\frac{2\gamma M^2 + \gamma +1}
                                   {\gamma +1}\,, 
      \\ \noalign{\medskip}
    v_{x,d} &=&\displaystyle \frac{v_{x,u}}{r} \,.
\end{array}\right.
\end{equation}
Here $\gamma=5/3$ is the adiabatic index, $M=v_{x,u}/c_{s, u}$ is the Mach number defined in terms of the sound speed $c_{s, u}=\sqrt{\gamma p_u/\rho_u}$, while 
\begin{equation}
    r=\frac{(\gamma + 1) M^2}{2 + (\gamma - 1)M^2}\simeq3.64
\end{equation}
is the shock compression ratio.
The magnetic field is imposed constant and parallel to the shock normal, that is $\mathbf{B} = (1,0,0)$.
We set periodic boundary conditions in the $y$ direction and outflow (zero-gradient) along $x$.
The MHD equations are solved using a second order Runge-Kutta time marching algorithm, linear reconstruction and the HLL Riemann solver \cite{harten1983}.
The divergence-free evolution of magnetic field is secured by using the upwind constrained transport method with the UCT\_HLL average \cite{Mignone2021}.

On top of the fluid we place $100$ particles in the upstream region, with energy spectrum initialized according to
%$i=1,...,N_{\rm bins}$, initializing  to a power--law ranging between $\epsilon_{\rm min}$ and $\epsilon_{\rm max}$ with slope $m$ according to
\begin{equation} \label{eq:initSpec}
    \chi_i=\mathcal{N}_m\Bigg(\frac{1-p}{\epsilon_{\rm max}^{1-p} - \epsilon_{\rm min}^{1-p}}\Bigg)\epsilon_i^{-p}\,,
\end{equation}
where $i=1,...,N_{\rm b}$ runs across the $N_{\rm b}=128$ energy bins between $\epsilon_{\rm min}=10^2$, $\epsilon_{\rm max}=10^4$, a power-law index $p=9$ and an initial fraction density of non-thermal to thermal particles $\mathcal{N}_0 =10^{-6}$.

\begin{figure}[h!]
    \centering
    \includegraphics[width=\columnwidth]{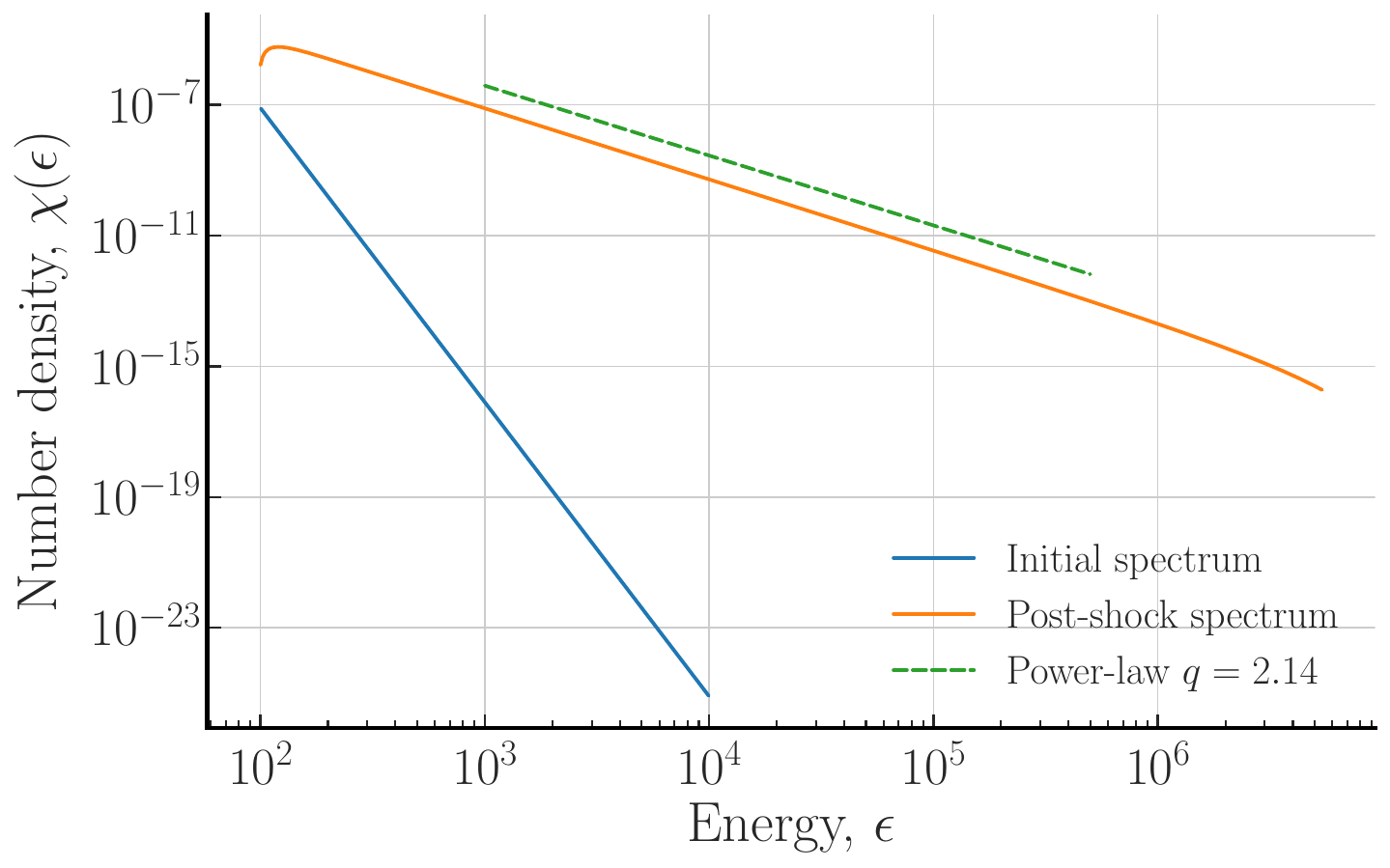}
    \caption{\footnotesize Initial (blue) and post--shock (orange) spectra of a LP that crosses the discontinuity. The green dashed line represents a power--law with index $q=2.14$ correspondent to the slope predicted by equation \ref{eq:slope} for an MHD shock with the compression ratio imposed in the present initial configuration. Note that the high--energy tail of the orange curve presents a deviation from a straight line as an effect of the synchrotron cooling induced by the magnetic field.}
    \label{fig:shock}
\end{figure}

In Figure \ref{fig:shock} we show the power spectrum of one LP before and after the shock crossing and, as a reference, a power-law with index $q(r=3.64)$ (see Eq. \ref{eq:slope}).
The slope of the reference power-law is in excellent agreement with the one of the post-shock spectrum, thereby demonstrating the correctness of the implementation on both the CPU and GPU.
{
The relative error between the slope of the power-law, $q_{\rm LP}$, and the theoretical slope $q(r=3.64)$ is $|q_{\rm LP} - q(r=3.64)|/q(r=3.64)=0.5\%$.
}

%%%%%%%%%%%%%%%%%%%%%%%%%%%%%%%%%%%%%%%%%%%%%%%%%%%%%%%
\subsection{Scalability}
\label{sec:performances}
%%%%%%%%%%%%%%%%%%%%%%%%%%%%%%%%%%%%%%%%%%%%%%%%%%%%%%%

Next, we present the scalability benchmarks conducted on the Euro-HPC pre-exascale supercomputers Marenostrum 5 and Leonardo hosted respectively at the Barcelona Supercomputing Center and CINECA.
The accelerated partition (ACC) of Marneostrum 5 is composed by 1120 nodes equipped with 2 Intel Sapphire Rapids 8460Y+ for a total of 80 cores and 4 64GB Nvidia Hopper GPU.
The general purpose partition (GCC) hosts 6408 nodes with 2 Intel Sapphire Rapids 8480+ for a total of 112 cores per node.
The Leonardo Booster partition presents 3456 nodes with 4 NVIDIA custom Ampere A100 GPU 64GB and one 32-core Intel Xeon Platinum 8358 CPU per node.
{
On both the machines we compiled using the recommanded software suite, e.g. nvc++ 23.11-0 for Marenostrum and nvc++ 24.5-1 for Leonardo.
}
We report in \ref{app2} further details on hardware and software specifications.

We employ both the advection and the MHD stationary shock setups described in Section \ref{sec:test} and extend the integration to 3D using 4 particles per grid cell.
In the simple advection setup, we prescribe the velocity field to be uniform and isotropic, $\textbf{v}=(0.2,0.2,0.2)$, removing the fluid freezing condition. We will refer to the two physical configurations as \textit{test $\mathcal{A}$} for the advection and \textit{test $\mathcal{S}$} for the shock.
For each CPU node we resolve $288\times288\times196\simeq1.6\times10^7$ grid points, or $N_{\rm p, CPU}\simeq6.8\times10^7$ particles, and we perform $200$ steps.
Given the 112 cores available in the CPU node, the grid is evenly distributed by using $4\times4\times7$ processors, each of which solves a sub-grid composed of $72\times72\times28$ points.
For the GPU partitions we use a grid of $320\times320\times160\simeq1.6\times10^7$ points, corresponding to $N_{\rm p, GPU}\simeq6.6\times10^7$ particles, and we evolve for $800$ steps.
In this case we subdivide the problem among $2\times2\times1$ GPUs resulting in $160\times160\times160$ cells per GPU (see Table \ref{tab:cpugpu}).
The physical domain of the node grid is defined as the interval $[-1,1]$ for the three integration directions.
The choice of such different grids depending on the partition is motivated by the necessity of employing in both cases all the CPUs/GPUs of the relative node, yet ensuring a even intra-node grid decomposition.
With these choices, the resolutions are such that the computational load is essentially the same on the two architectures: $\frac{N_{\rm p, CPU}}{N_{\rm p, GPU}}\sim 0.99$.
{
%We decided upon fixing the steps number, rather than the final simulation time, to ensure that every run would perform the same amount of operations, independently on the time step size, which varies with the cell size.
Note that when deciding the number of steps for the CPU and GPU cases, we balanced the benefits of having long tests (e.g. suppressing initialization effects) with the a finite amount of computational hours.
The numeric discrepancy that arise from the inherently higher GPU performance will be factorized out by normalizing the results accordingly.}
The benchmark problem sizes have been tuned to obtain a GPU memory usage (measured with \texttt{nvidia-smi}) of $\sim90\%$ to maximize GPU load yet avoiding saturation.
Note that we chose the ratio between the grid size and the number of particles so that the computational load is dominated by the particle update rather than by the fluid.
This is made clear in Table \ref{tab:codeTiming} where, for the test $\mathcal{A}$ evaluated on a full node of Leonardo Booster, the particle routines take $85\%$ of the total computational time.
%We consider reasonable to assume this result to hold also for configuration $\mathcal{S}$ and for all the partitions used in the present work.
%
\begin{table}[!ht]
\centering
\resizebox{\linewidth}{!}{  % Scale to column width
\begin{tabular}{@{}lccccc@{}}
\toprule
 &
  \begin{tabular}[c]{@{}c@{}}No. of\\cores\end{tabular} &
  \begin{tabular}[c]{@{}c@{}}Node\\gird\end{tabular}&
  \begin{tabular}[c]{@{}c@{}}No. of cores\\per direction\end{tabular} &
   \begin{tabular}[c]{@{}c@{}}Core\\grid\end{tabular} & \begin{tabular}[c]{@{}c@{}}No. of\\
   steps\end{tabular}  \\ \midrule
CPU &
  112 &
  $288^2\times196$ &
  $4^2\times7$ &
  $72^2\times28$ &200\\
GPU &
  4 &
  $320^2\times160$ &
  $2^2\times1$ &
  $160^3$&800 \\\midrule\bottomrule
\end{tabular}
}
\caption{The grid configurations utilized for the CPU and GPU scalability tests.}
\label{tab:cpugpu}
\end{table}

\paragraph{Strong scaling}

Here we measure the time to solution as a function of the number of processors while keeping the base problem size fixed.
If $n$ is the number of nodes used in the computation, the speedup is defined as the ratio $T^\mathcal{X}_1/T^\mathcal{X}_n$ where $T^\mathcal{X}_1$ and $T^\mathcal{X}_n$ are the application execution times on $1$ and $n$ nodes, respectively, and the superscript $\mathcal{X}=\{\mathcal{A}, \mathcal{S}\}$ refers to the advection or the shock test problems.

\begin{figure}[!ht]
    \centering
    \includegraphics[width=\columnwidth]{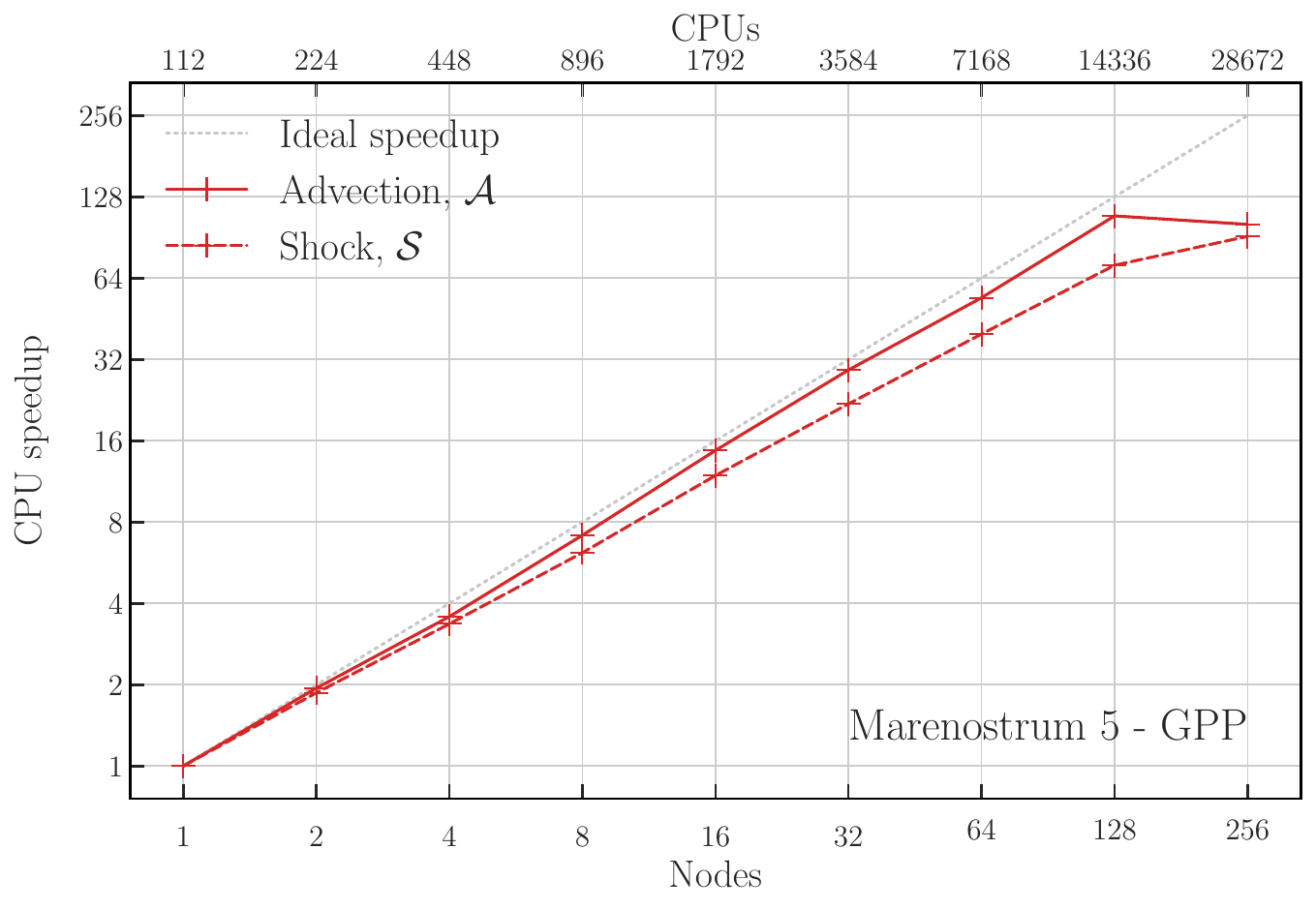}
    \caption{\footnotesize CPU speedup on Marenostrum 5 GPP partition. The red and orange solid lines represent the measured speedup for the advection and the shock test problem respectively. The grey dotted line is the theoretical speedup correspondent to a linear inverse relation of the time to solution with the number of cores employed.}
    \label{fig:strongCPU}
\end{figure}

Figure \ref{fig:strongCPU} shows the application speedup as a function of the number of CPU nodes utilized for the two test cases.
The relative times to solution $T_n^{\mathcal{A}}$ and $T_n^{\mathcal{S}}$ are reported in Table \ref{tab:cpuStrong}.
\begin{table}[!ht]
\centering
%\resizebox{\linewidth}{!}{  % Scale to column width
%\scalebox{0.8}{  % Scale to column width
\begin{tabular}{@{}c|ccccccccc@{}}
\toprule
\textbf{Nodes, $n$}    & 1   & 2   & 4   & 8  & 16 & 32 & 64 & 128 & 256 \\\midrule
$T^{\mathcal{A}}_n$ [s] & 348 & 179 & 97  & 48 & 21 & 11 & 6  & 3   & 3   \\
$T^{\mathcal{S}}_n$ [s] & 364 & 195 & 108 & 59 & 26 & 16 & 9  & 5   & 3  \\
\midrule\bottomrule
\end{tabular}
\caption{Number of nodes and execution times in seconds of of the two test problems for the CPU strong scaling on Marenostrum 5. The same data are plotted in form of speedup in Figure \ref{fig:strongCPU}.}
\label{tab:cpuStrong}
\end{table}
Test $\mathcal{A}$ demonstrates nearly-ideal scaling up to $128$ nodes ($14336$ MPI tasks) and a net breakdown for 256 nodes, where the computational load becomes too small to provide further performance increase.
On the other hand, configuration $\mathcal{S}$ starts deviating from the ideal behaviour at $32$ nodes.

When switching to GPUs, it is important to keep in mind that computational efficiency is fully exploited when occupancy is maximized. 
This translates in a GPU speedup that deviates from the theoretical behaviour far earlier than what happens for the CPUs.
In order to measure strong scaling capabilities up to $128$ GPU, we introduce five different speedup curves defined as the ratio $T^\mathcal{X}_{b}/T^\mathcal{X}_n$ where $T^\mathcal{X}_b$ is the execution time of the chosen test problem using $b\in\{1,2,4,8,16\}$ base nodes and $n\geq b$.
In particular, for each doubling of $b$, the grid is also doubled in one direction as shown in Table \ref{tab:gpustrong}.
We also modify the physical boundary according to the grid size so to keep the cell size constant.
\begin{table}[!ht]
\centering
\begin{tabular}{@{}lcccc@{}}
\toprule
\multicolumn{1}{c|}{}        & \multicolumn{2}{c}{Leonardo Booster}       & \multicolumn{2}{c}{Marenostrum ACC}       \\ \midrule
\multicolumn{1}{c|}{\textbf{Nodes, $n$}} & $T_n^{\mathcal{A}}$ [s] & $T_n^{\mathcal{S}}$ [s] & $T_n^{\mathcal{A}}$ [s] & $T_n^{\mathcal{S}}$ [s] \\ \midrule
\multicolumn{5}{c}{\textbf{$\mathbf{b=1}$,\quad Grid: $\mathbf{320^2\times160}$,\quad $\mathbf{N_p=6.6\times10^7}$}}  \\ \midrule
\multicolumn{1}{l|}{1}       & 361                  & 544                 & 215                 & 401                 \\
\multicolumn{1}{l|}{2}      & 176                  & 253                 & 124                 & 184                 \\
\multicolumn{1}{l|}{4}       & 96                   & 121                 & 79                  & 94                  \\
\multicolumn{1}{l|}{8}       & 61                   & 83                  & 55                  & 66                  \\ \midrule
\multicolumn{5}{c}{\textbf{$\mathbf{b=2}$,\quad Grid: $\mathbf{320^2\times320}$,\quad $\mathbf{N_p=1.3\times10^8}$}}  \\ \midrule
\multicolumn{1}{l|}{2}       & 371                  & 562                 & 235                 & 421                 \\
\multicolumn{1}{l|}{4}       & 177                  & 249                 & 129                 & 189                 \\
\multicolumn{1}{l|}{8}       & 100                  & 128                 & 84                  & 97                  \\
\multicolumn{1}{l|}{16}      & 64                   & 88                  & 57                  & 73                  \\ \midrule
\multicolumn{5}{c}{\textbf{$\mathbf{b=4}$,\quad Grid: $\mathbf{640\times320^2}$,\quad $\mathbf{N_p=2.6\times10^8}$}}  \\ \midrule
\multicolumn{1}{l|}{4}       & 382                  & 564                 & 240                 & 422                 \\
\multicolumn{1}{l|}{8}       & 178                  & 257                 & 134                 & 190                 \\
\multicolumn{1}{l|}{16}      & 103                  & 130                 & 87                  & 103                 \\
\multicolumn{1}{l|}{32}      & 65                   & 88                  & 59                  & 78                  \\ \midrule
\multicolumn{5}{c}{\textbf{$\mathbf{b=8}$,\quad Grid: $\mathbf{640^2\times320}$,\quad $\mathbf{N_p=5.2\times10^8}$}}  \\ \midrule
\multicolumn{1}{l|}{8}       & 387                  & 567                 & 246                 & 424                 \\
\multicolumn{1}{l|}{16}      & 181                  & 267                 & 137                 & 196                 \\
\multicolumn{1}{l|}{32}      & 107                  & 135                 & 88                  & 105                 \\
\multicolumn{1}{l|}{64}      & 76                   & 103                 & 60                  & 76                  \\ \midrule
\multicolumn{5}{c}{\textbf{$\mathbf{b=16}$,\quad Grid: $\mathbf{640^2\times640}$,\quad $\mathbf{N_p=1.0\times10^9}$}} \\ \midrule
\multicolumn{1}{l|}{16}      & 461                  & 634                 & 253                 & 433                 \\
\multicolumn{1}{l|}{32}      & 190                  & 275                 & 138                 & 197                 \\
\multicolumn{1}{l|}{64}      & 120                  & 150                 & 91                  & 110                 \\
\multicolumn{1}{l|}{128}     & 137                  & 157                 & 61                  & 86                  \\ \bottomrule
\bottomrule
\end{tabular}
\caption{Number of nodes and execution times in seconds of both configuration $\mathcal{A}$ $\mathcal{S}$ for the five grid base configurations. The timings are obtained on the Booster partition of Leonardo.}
\label{tab:gpustrong}
\end{table}

\begin{figure*}[!ht]
    \centering
    \includegraphics[width=\textwidth]{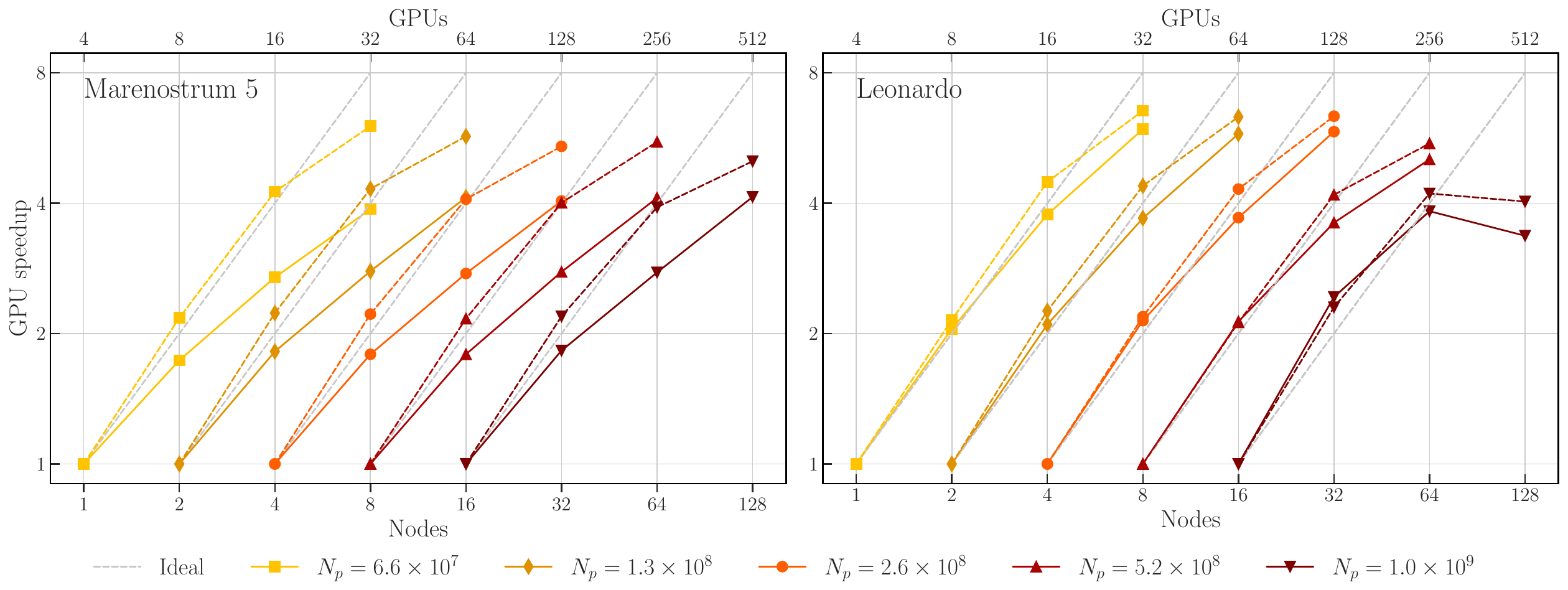}
    \caption{GPU strong scaling speedup on { Marenostrum 5 (left panel)} and Leonardo Booster {(right panel)}. Each pair of curves with the same marker and the same colour refers to a given problem size of Table \ref{tab:gpustrong}. The solid lines refer to problem $\mathcal{A}$ whereas the dashed lines are related to the problem $\mathcal{S}$. The grey dotted line is the theoretical speedup.}
    \label{fig:strongGPUboth}
\end{figure*}

Figure \ref{fig:strongGPUboth} shows the speedup evaluated on the ACC partition of Marenostrum 5 { (left panel) and the Booster partition of Leonardo (right panel).}
Each pair of lines (solid and dashed with the same marker) corresponds to a fixed problem size.
The solid line represent test $\mathcal{A}$, while the dashed one refers to test $\mathcal{S}$.
{
%For the run on Marenostrum case $\mathcal{A}$ shows an immediate breakdown of the strong scaling speedup.
%On the other hand both the tests on Leonardo and test $\mathcal{S}$ on Marenostrum deviate from the theoretical speedup line after two node doubling independently on grid size.
Here, while configuration $\mathcal{S}$ shows good strong scaling speedup up to $N_{\rm p, node} \lesssim N_{\rm p}/4$ on both clusters, configuration $\mathcal{A}$ experiences a significant performance breakdown immediately following the first node increase on Marenostrum 5.
This comes from $\mathcal{S}$ presenting higher computational cost because of the post-shock particles spectral update.
Moreover, only a subset of the particles to experience the post-shock update, this forces threads to diverge and worsen overall kernel performances in test $\mathcal{S}$.
The reduced time-to-solution of test $\mathcal{A}$   with similar communication overheads (see also Table \ref{tab:gpustrong}) makes the latter relatively more important with respect to the overall execution time, leading to worse scaling performances.
The behaviour is more evident when using more powerful GPUs, such as the H100 Nvidia GPU available in Marenostrum 5.

Both panels in Fig. \ref{fig:strongGPUboth} show a modest super-linear trend, whereby doubling the number of nodes results in a slightly higher speedup than the ideal linear estimate.
This is likely due to a predominantly memory-bounded application, meaning that reducing the per-GPU problem size allows to reduce register and cache pressure, leading to slightly increased computational throughput.
As shown in Table \ref{tab:gpustrong}, the deviation from ideal strong scaling remains limited, with
\begin{equation}
    \frac{|T_{\rm th} - T_{\rm 2b}|}{T_{\rm th}}\lesssim10\%\;,
\end{equation}
where $T_{\rm th}=T_{\rm b}/2$.
Therefore, we acknowledge that maximizing GPU memory utilization does not always coincide with optimal performance, especially in memory-bound applications where memory pressure can limit throughput, however the chosen configuration, with approximately $90\%$ memory occupancy, remains close to the optimal regime and is adequate for the purposes of the present study.
}

\paragraph{Weak scaling}
We now turn our attention to the parallel efficiency, or the time to solution variation $T^\mathcal{X}_1/T^\mathcal{X}_n$ as the number of processors increases linearly with the problem size.
For each node increment we increase the grid and the domain boundaries according to what reported in Table \ref{tab:gridWeak}.
\begin{table}[!ht]
\resizebox{\linewidth}{!}{  % Scale to column width
\begin{tabular}{@{}cccc@{}}
\toprule
\multicolumn{1}{c|}{\textbf{Nodes, $n$}} & \textbf{CPU Grid} & \textbf{GPU Grid} & \textbf{Boundary} \\ \midrule
\multicolumn{1}{c|}{1}   & $288\times288\times196$ & $320^2\times160$ & $[-1,1]^2\times[-1,1]$ \\
\multicolumn{1}{c|}{2}   & $576\times288\times392$ & $320^2\times320$ & $[-1,1]^2\times[-2,2]$ \\
\multicolumn{1}{c|}{4}   &  $576\times576\times392$ & $640\times320^2$ & $[-2,2]\times[-1,1]^2$ \\
\multicolumn{1}{c|}{8}   &  $576\times576\times784$ & $640^2\times320$ & $[-2,2]^2\times[-1,1]$ \\
\multicolumn{1}{c|}{16}  & $1152\times576\times784$ & $640^2\times640$ & $[-2,2]^2\times[-2,2]$ \\
\multicolumn{1}{c|}{32}  &  $1152\times1152\times784$ & $1280\times640^2$ & $[-4,4]\times[-2,2]^2$ \\
\multicolumn{1}{c|}{64}  & $1152\times1152\times1568$ & $1280^2\times640$ & $[-4,4]^2\times[-2,2]$ \\
\multicolumn{1}{c|}{128} & $2304\times1152\times1568$ & $1280^2\times1280$ & $[-4,4]^2\times[-4,4]$ \\
\multicolumn{1}{c|}{256} & $2304\times2304\times1568$ & $2560\times1280^2$ & $[-8,8]\times[-4,4]^2$ \\ \midrule\bottomrule
\end{tabular}
}
\caption{Grid size and physical boundaries as a function of the number of nodes utilized in the weak scaling tests. The grid decomposition is different for the CPU and GPU partitions.}
\label{tab:gridWeak}
\end{table}

\begin{figure}[!ht]
    \centering
    \includegraphics[width=\columnwidth]{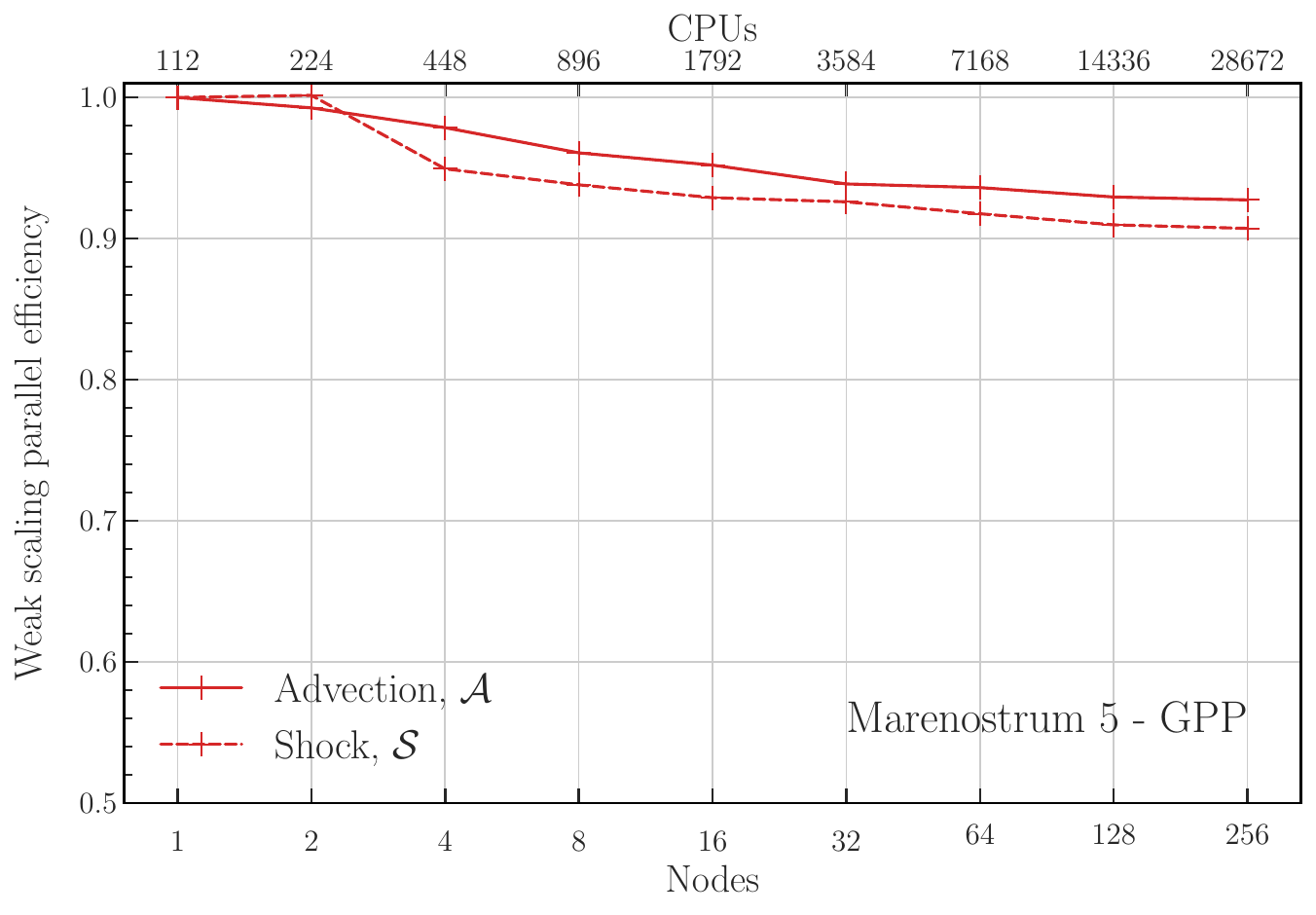}
    \caption{\footnotesize Weak scaling parallel efficiency up to 256 nodes of test $\mathcal{A}$ the advection (red line) and test $\mathcal{S}$ (orange line) measured on the GPP partition of Marenostrum 5.}
    \label{fig:weakCPU}
\end{figure}

\begin{figure}[!ht]
    \centering
    \includegraphics[width=\columnwidth]{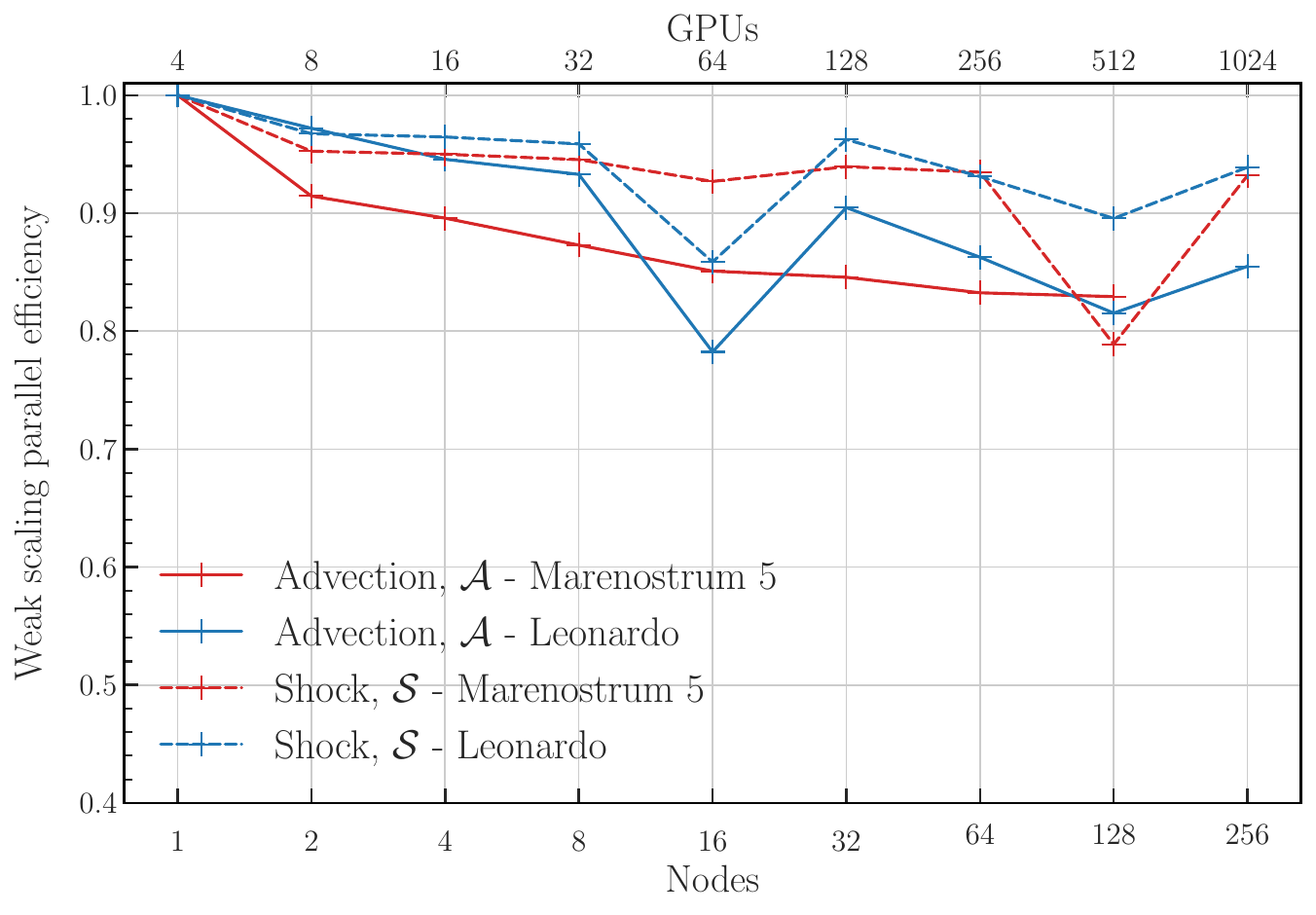}
    \caption{\footnotesize Weak scaling parallel efficiency for the advection (solid lines) and the shock (dashed lines) configurations. The red lines refer to Leonardo whereas the blue lines are related to Marenostrum 5.}
    \label{fig:weakGPU}
\end{figure}

Figure \ref{fig:weakCPU} shows that the parallel efficiency of both the test problems remains above $90\%$ up to 256 node or 28672 MPI processes on the GPP partition of Marenostrum 5.
In Figure \ref{fig:weakGPU} we measure the code parallel efficiency on the ACC partition of Marenostrum 5 and the Booster partition of Leonardo.
Configuration $\mathcal{S}$ shows a parallel efficiency of $90\%$ up to 256 nodes or 1024 GPUs on both machines yet presenting dips of performances at 16 nodes in Leonardo and at 128 nodes in both the clusters.
Test case $\mathcal{A}$ demonstrates a parallel efficiency higher than $80\%$ and presents the same performances dips at $16$ and $128$ nodes in the Leonardo runs while no particular out-of-trend data points are observed for Marenostrum 5.
We point out that we could not perform the $256$ nodes run on Marenostrum 5 due to limited hours budget on the cluster. 

Table \ref{tab:weak} reports the time to solution of the weak scalings as a function of the number of nodes for the advection and the shock test problems evaluated on Marenostrum 5 GPP and ACC and on Leonardo Booster.
\begin{table}[!ht]
\centering
\resizebox{\linewidth}{!}{  % Scale to column width
%\scalebox{0.8}{  % Scale to column width
\begin{tabular}{@{}ccccccc@{}}
\toprule
 &
  \multicolumn{2}{c}{\begin{tabular}[c]{@{}c@{}}Marenostrum GPP\\No. of steps 200\end{tabular}} &
  \multicolumn{2}{c}{\begin{tabular}[c]{@{}c@{}}Marenostrum ACC\\No. of steps 800\end{tabular}} &
  \multicolumn{2}{c}{\begin{tabular}[c]{@{}c@{}}Leonardo Booster\\No. of steps 800\end{tabular}} \\ \midrule
\multicolumn{1}{c|}{\small\textbf{Nodes, $n$}} &
  $T_n^{\mathcal{A}}$ [s] &
  $T_n^{\mathcal{S}}$ [s] &
  $T_n^{\mathcal{A}}$ [s] &
  $T_n^{\mathcal{S}}$ [s] &
  $T_n^{\mathcal{A}}$ [s] &
  $T_n^{\mathcal{S}}$ [s] \\ \midrule
\multicolumn{1}{c|}{1}   & 348 & 364 &  215 & 401 & 361 & 544 \\
\multicolumn{1}{c|}{2}   & 350 & 363 &  235 & 421 & 371 & 562 \\
\multicolumn{1}{c|}{4}   & 355 & 383 &  240 & 422 & 382 & 564  \\
\multicolumn{1}{c|}{8}   & 362 & 388 &  246 & 424 & 387 & 567  \\
\multicolumn{1}{c|}{16}  & 365 & 392 &  253 & 433 & 461 & 634  \\
\multicolumn{1}{c|}{32}  & 371 & 393 &  254 & 427 & 399 & 565  \\
\multicolumn{1}{c|}{64}  & 372 & 397 &  258 & 429 & 418 & 584   \\
\multicolumn{1}{c|}{128} & 374 & 400 &  260 & 509 & 443 & 608   \\
\multicolumn{1}{c|}{256} & 375 & 401 & - & 431 & 422 & 579   \\ \midrule\bottomrule
\end{tabular}
}
\caption{Time to solution in seconds as a function of the number of nodes for the test $\mathcal{A}$ and $\mathcal{S}$ evaluated on the three different partitions. Note that the number of steps is different for CPU and GPU.}
\label{tab:weak}
\end{table}

{
\paragraph{Cluster comparison}
We compare here the GPU code performances on A100 Leonardo's GPUs against H100 chips of MArenostrum 5.
Figure \ref{fig:ClusterComp} shows the ratio $T^\mathcal{X}_{n,{\rm Leonardo}}/T^\mathcal{X}_{n,{\rm Marenostrum}}$ of the execution times reported in Table \ref{tab:weak}.
According to the results, the architecture of Marensotrum 5 grants $\sim(40-80)\%$ more performance with respect to Leonardo Booster.
The peak at 16 nodes corresponds to an increased time-to-solution of Leonardo booster, also visible in Figure \ref{fig:weakGPU}.
\begin{figure}[!ht]
    \centering
    \includegraphics[width=\columnwidth]{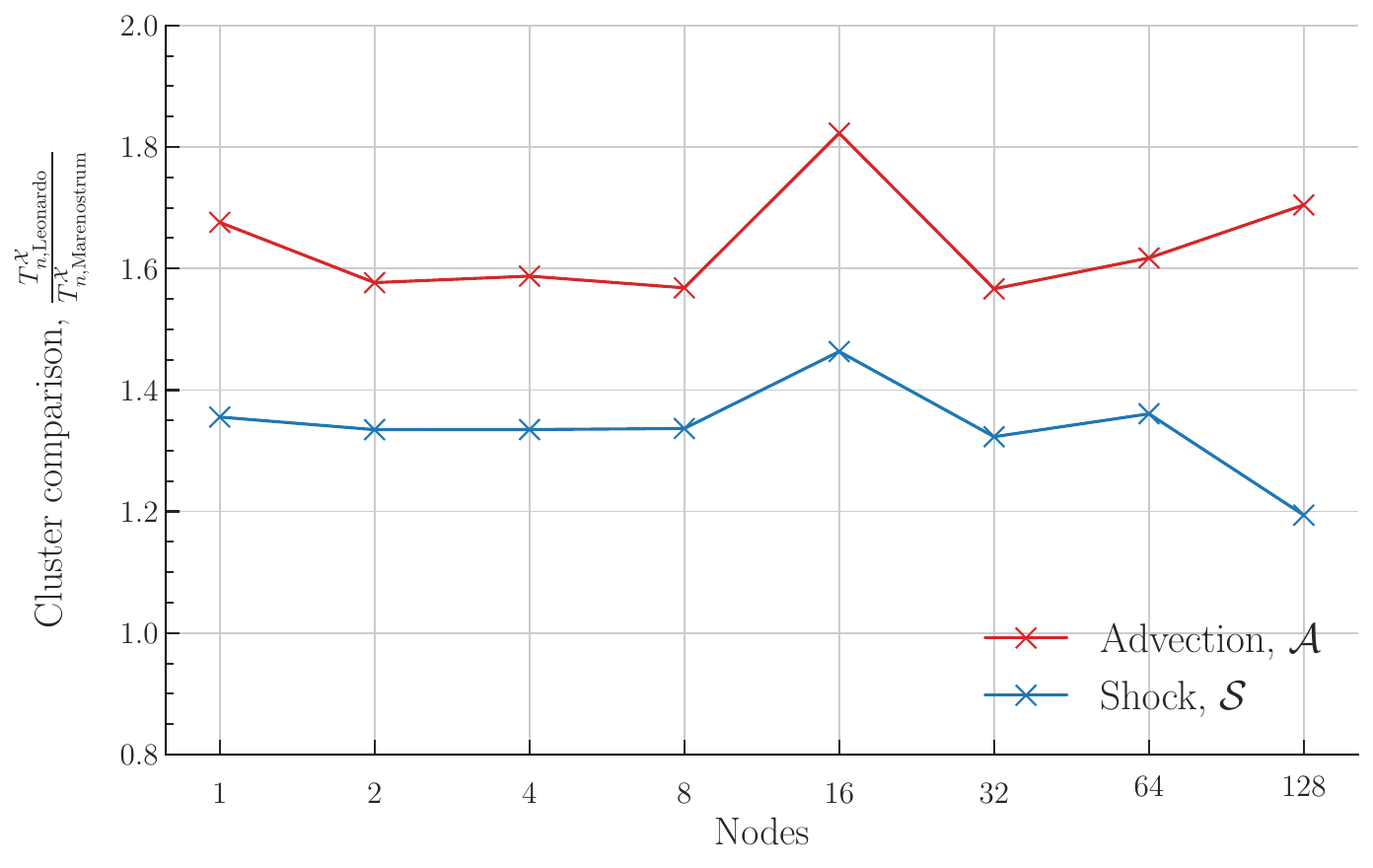}
    \caption{\footnotesize  Time to solution comparison of the same GPU configuration as a function of the number of nodes between Marenostrum 5 and Leonardo.}
    \label{fig:ClusterComp}
\end{figure}

}

\begin{figure}[!ht]
    \centering
    \includegraphics[width=\columnwidth]{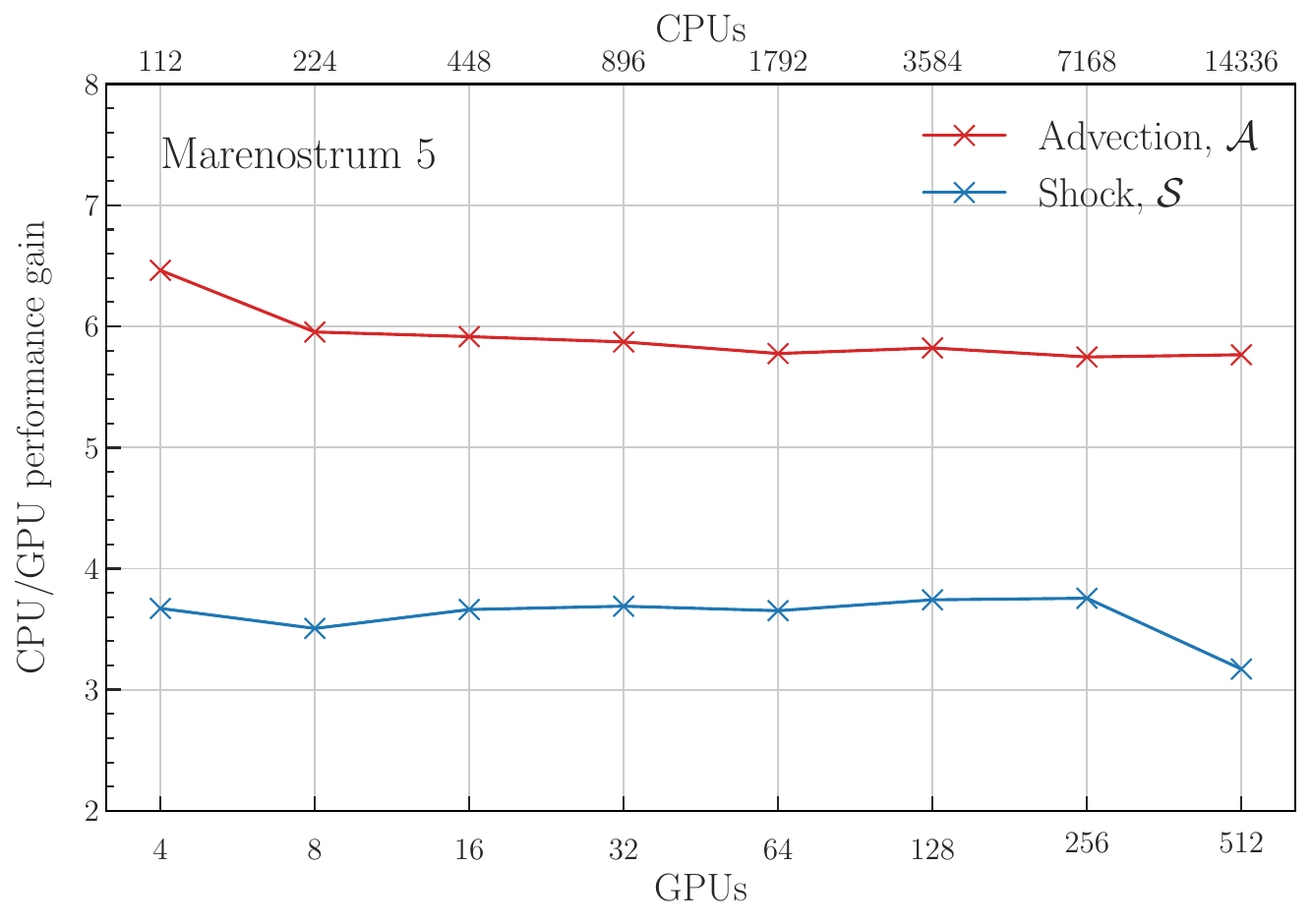}
    \caption{\footnotesize Performance gained by running { \texttt{gPLUTO}} on the ACC partition of Marenostrum 5 with respect to the GPP partition as a function of the number of nodes. The red lines refers to test $\mathcal{A}$, whereas the blue line is related to test $\mathcal{S}$. { The problem size is increased as a function of the number of nodes according to Table \ref{tab:gridWeak}.}}
    \label{fig:CPU-GPUspeedup}
\end{figure}

\paragraph{GPU/CPU acceleration}
We measure here the performance gain $T^\mathcal{X}_{n,{\rm CPU}}/T^\mathcal{X}_{n,{\rm GPU}}$ given by a full GPU node(s) run on the ACC partition in Marenostrum 5 with respect a full CPU node(s) of the GPP partition.
Figure \ref{fig:CPU-GPUspeedup} shows a performance gain between $3\textrm{ and }4$ up to 128 nodes for the shock configuration, which becomes $6$ for the advection case.

\section{Discussion and conclusions}
\label{sec:conclusion}
In the present work, we have described a revised implementation of the Lagrangian Particles module of the PLUTO code \cite{vaidya2018}, more suitable for modern (pre-)exascale computing architectures. 
The module predicts the non-thermal spectral evolution of NTPs through a variety of mechanisms such as synchrotron and inverse Compton emission, adiabatic heating and expansion, as well as Fermi I acceleration at MHD shocks.

The code employs the OpenACC programming model to accelerate computations through offloading to NVIDIA GPUs, while maintaining CPU compatibility and code simplicity.
We outlined the fundamental concepts behind GPUs architecture and operation and summarized some programming best practises to better utilize their computing power.
We focused our analysis on two numerical benchmarks with analytical reference solutions to validate the newly designed implementation and assess the accuracy of the GPU results against theoretical predictions.
Moreover, the scaling capabilities of the code on two different Euro-HPC clusters (up to 28672 CPUs and 1024 GPUs) have been measured.

{
Strong scaling performance across the CPU and GPU partitions have been presented in Figures \ref{fig:strongCPU} and \ref{fig:strongGPUboth}.
The implementation achieves nearly ideal speedup on the MareNostrum 5 GPP (CPU) partition up to 128 nodes demonstrating an effective domain decomposition and low communication overhead for traditional architectures.
On GPU, we observe a robust strong scaling up to 4 nodes (16 GPUs) for both configurations on Leonardo Booster and on the Shock test on Marenostrum 5, whereas the advection test shows a sub-linear speedup already from the first node increase.
The early breakdown of the speedup curve for the GPU partitions (in comparison from the very late speedup breakdown of the the CPU case) reflects the fact that the GPUs' massive parallelization capabilities is fully exploited only when computation is performed on large amount of data  and threads can efficiently hide latencies.

Furthermore, we notice that on CPUs the shock test exhibits slightly higher deviation from ideal scaling than advection, and that the situation is reversed in the GPU case.
This is because CPU is affected by the load unbalance introduced in shock-related operations, leading to a reduced parallel efficiency, whereas for GPUs, an higher algorithmic complexity increases computation load and mitigates the relative impact of the inter-core communication.
%resulting in better strong scaling compared to the advection case.
This behaviour becomes even more evident when the advection test is benchmarked on MareNostrum 5 ACC, which is equipped with more powerful GPUs with respect to Leonardo (see Table \ref{tab:clusters} and the cluster comparison in Figure \ref{fig:ClusterComp}).

%For these tests the time-to-solution reduction given by solving the same problem with more GPUs becomes sub-liner when the number of particles per node $N_{\rm p, node} \lesssim N_{\rm p}/4$. % (see figure \ref{fig:strongGPUboth}).
% The discrepancy between the CPU and the GPU strong scaling behaviours comes from their inherently operational paradigms.
% Higher GPU occupancy allows it to increase the number of operations performed in parallel whereas a CPU works sequentially no-matter the size of the problem which translates in an early breakdown of the strong scaling speedup curve.
}
% In fact, in contrast on what happens for CPUs, the fact that GPUs are able to parallelize computation translate to a sub-linear reduction of the time-to-solution when incrementing the number of cores utilized to solve a given problem.

Regarding the weak scalings, the code showed a parallel efficiency that, after a first decrease up to $4$ nodes, stabilizes around $90\%$ for the CPUs { (Figure \ref{fig:weakCPU})} and $(80-90)\%$ for GPUs { (Figure \ref{fig:weakGPU})}.
{
We also notice how the advection configuration shows better scaling properties on all machines, supporting the idea that higher computational throughput at kernel level negatively impacts the scaling capability given a fixed communication performance.
}
While the parallel efficiency curves on CPU show a smooth trend, we notice significant dips at { 16} and 128 nodes for the case of the GPUs.
These features may be caused by the code itself, the clusters architecture, the simulation setup or a combination of these and other unknown factors.
We commit to pose our future attention in understating such inefficiencies in a joint effort with HPC and GPU experts.

Finally we assessed the performance gained by solving the same problem with an accelerated partition equipped with 4 GPUs per node against a CPU partition with 112 CPUs per node.
The GPU execution is $\sim4$ times faster on GPU for the case of the shock setup and $\sim6$ times faster for the advection { (Figure \ref{fig:CPU-GPUspeedup})}.
{
We notice that, even though the scaling tests look more favourable for the shock configuration, the overall performance gain is significantly higher for the advection one at any number of nodes.
}
We commit to address the remaining inefficiency, especially for load-unbalanced problems such as the shock configuration, maximizing the level of exposed parallelism in future dedicated optimizations cycles.

At last we remark that 
%Future works will also explore additional acceleration frameworks such as OpenMP for execution on AMD or Intel based accelerators.
the acceleration model of choice, OpenACC, supports NVIDIA GPUs acceleration only, limiting the code from exploiting other vendors devices (such as AMD or Intel).
Even though this issue is being currently addressed through the implementation of OpenMP, a similar and yet more portable programming model, it is in its early stages and we leave its discussion for future works.

\section{Acknowledgements}
AS thanks M. Bettencourt from NVIDIA for his competence, his patience and his active support in the early stages of the porting here described. AS also thanks M. Mencagli for HPC support on Leonardo and consulting on kernel optimization.
{A mention goes to the profound questions posed by the referees which helped improving the present manuscript.}

The authors acknowledge ISCRA for awarding this project access to the Leonardo supercomputer and the EuroHPC Joint Undertaking for granting us access to the Leonardo and MareNostrum 5 supercomputers through an EuroHPC [Extreme/Regular/Benchmark/Development/. . . ] Access call.

This paper is supported by the Fondazione ICSC, Spoke 1 "FutureHPC \& BigData” and Spoke 3 Astrophysics and Cosmos Observations and National Recovery and Resilience Plan (Piano Nazionale di Ripresa e Resilienza, PNRR), Project ID CN 00000013 “Italian Research Center on High-Performance Computing, Big Data and Quantum Computing” funded by MUR Missione 4 Componente 2 Investimento 1.4: Potenziamento strutture di ricerca e creazione di “campioni nazionali di R\&S (M4C2-19)” - Next Generation EU (NGEU).

% This work has been supported by the Spoke-1 "FutureHPC \& BigData” of the ICSC – Centro Nazionale di Ricerca in High Performance Computing, Big Data and Quantum Computing – and hosting entity, funded by European Union – NextGenerationEU.

This work has also received funding from the European High Performance Computing Joint Undertaking (JU) and Belgium, Czech Republic, France, Germany, Greece, Italy, Norway, and Spain under grant agreement No 101093441 (SPACE).

\bibliographystyle{elsarticle-num}
\bibliography{biblio}

\appendix
\section{Containers for particle allocation}
\label{sec:containers}
We report here the core implementation of the Particles Container structure.
As described in section \ref{sec:mem-manag} we decide to store the particles in a structure of arrays, where every array corresponds to a characteristic of the particle.
We define a templated class that can handle chunked memory allocation/deallocation which can be compile time instantiated to any C++ object. 
%Here the main methods and data members of the BlockedContainer:
\begin{lstlisting}[
    caption=BlockedContainer,
    label={lst:BC},
    style=vscodewhite
    ]
template <class T, unsigned int e_ = 16, unsigned int N_ = 0>
class BlockedContainer {
private:
  std::vector<T*> data_;
  constexpr static int L_ = 1 << e_;
  int N_ = 0;  /* Number of elements */
  int Nc_ = 0; /* Number of chunks   */
public:
  void Reserve(const unsigned int nNew)
  {
    N_ += nNew;
    const int Nc_new = (N_ >> e_)
        + (N_ % L_ == 0 ? 0 : 1 );
    for(int i = Nc_; i < (Nc_new); i++){
      data_.push_back(new T[L_]);
    }
    Nc_ = Nc_new;
    #pragma acc update device(this)
  }

  Shrink(const unsigned int nDel){
    N_ -= nDel;
    const int Nc_new = (N_ >> e_)
        + (N_ % L_ == 0 ? 0 : 1 );
     for(int i = Nc_new; i < Nc_; i++){
      delete [] data_.back();
      data_.pop_back();
    }
    Nc_ = Nc_new;
    #pragma acc update device(this)
  }

  inline int Size() const  {
    return N_;
  }

  T &operator() (const int i) const {
    return 
      (ptr_[i_ >> e_])
      [i & (L_ - 1)];
  }
};
\end{lstlisting}
The class BlockedContainer (BC) contains four private data members:
\begin{itemize}
    \item an \texttt{std::vector<T*>} containing the pointers to the data chunks of type T;
    \item the length of the chunk L\_, evaluated compile time thanks to the qualifier constexpr;
    \item the total number of element N\_ stored in the structure. Note that in the context of the present implementation we refer to this quantity with the attribute \textit{size};
    \item the number of data chunks Nc\_;
\end{itemize}
and is equipped with the following functions:
\begin{itemize}
    \item Reserve() increases the internal capacity to accommodate nNew additional elements. It evaluates the number of memory chunks needed, it allocates them dynamically and it appends them in data\_. Finally, the updated object is synchronized to the device.
    \item Shrink() decreases the capacity by deleting the nDel trailing items and takes care of chunks deallocation. It also synchronizes the device copy.
    \item Size() returns the number of element present in the structures 
    \item The function call operator () is overloaded to define a unique access pattern to the element i, hiding the chunked nature of the structure.
\end{itemize}
Note that in general the size (or the number of elements in  a BC) N\_ does not correspond to the total number of memory spaces allocated: \ N\_ $\leq$ (L\_ Nc\_).

For every core it is allocated one Particle Container (PC) that contains all the particles of the rank. The object consists of a set of BCs instantiated to the desired types that define a particle (for instance 3 double for the position, 3 double for the speed, 128 double for the energy spectrum, one 64-bit unsigned integer for the identification number, etc).
\begin{lstlisting}[
    caption=ParticlesContainer,
    label={lst:PC},
    style=vscodewhite
    ]
class ParticlesContainer {
private:
  constexpr int e_ = 18;
  using Coor   std::array <double,   3>;
  using Nebins std::vector<double, 128>;
  BlockedContainer<Coor    , e_> pos_;
  BlockedContainer<Coor    , e_> speed_;
  BlockedContainer<Nebins  , e_> chi_;
  BlockedContainer<Nebins  , e_> eng_;
  BlockedContainer<uint64_t, e_> id_;

public:
  void Reserve(const int nNew)
  {
    pos_.  Reserve(nNew);
    speed_.Reserve(nNew);
    ...
    #pragma acc update device(this)
  }

  void Shrink(const int nDel)
  {
    pos_.  Shrink(nDel);
    speed_.Shrink(nDel);
    ...
    #pragma acc update device(this)
  }

  inline int Size() const {
    return pos_.Size();
  }
  
  #pragma acc routine seq
  Coor &pos (const int i) {
    return pos_(i);
  }

  #pragma acc routine seq
  Coor &speed (const int i) {
    return speed_(i);
  }

  #pragma acc routine seq
  Nebins  &chi (const int i) {
    return chi_(i);
  }
  
  #pragma acc routine seq
  Nebins &eng (const int i) {
    return eng_(i);
  }

  #pragma acc routine seq
  uint64_t &id (const int i) {
    return id_(i);
  }
};
\end{lstlisting}
On top the class is equipped with the following helper functions
\begin{itemize}
    \item Reserve() allocates nNew particles by increasing the size of all the BCs that define the particle.  
    \item Shrink() deletes nDel trailing particles by shrinking all BCs. 
    \item Size() returns the total number particles currently allocated in the PC by returning the number of elements N\_ of one of the BCs. Note that all the BCs stored in the PC have the same size at all times, which corresponds to the number of allocated particles.
    \item pos() wraps the function call operato () of the corresponding BC to guarantee access to the position of the $i$-th particle. For every BC, a wrapper to the corresponding operator is provided.
\end{itemize}

\section{Hardware and software specifications}
\label{app2}
Table \ref{tab:clusters} summarizes the relevant technical data for the clusters utilized.

% \usepackage{booktabs}
\begin{table*}[!ht]
\begin{tabular}{@{}l|cc|c@{}}
\toprule
\textbf{}                                    & \multicolumn{2}{c|}{\textbf{Marenostrum}}                  & \textbf{Leonardo}         \\
                     & \multicolumn{2}{c|}{Barcelona Supercomputing Center (BSC)} & CINECA                    \\\midrule
\textbf{Partition Name}                      & \multicolumn{1}{c|}{ACC}                 & GPP             & Booster                   \\
\textbf{No. of nodes}                             & \multicolumn{1}{c|}{1120}                & 6408            & 3456                      \\
\textbf{CPU model} & \multicolumn{1}{c|}{Intel Sapphire Rapids 8460Y+} & Intel Sapphire Rapids 8480+ & Intel Xeon Platinum 8358 \\
\textbf{No. of sockets $\mathbf{\times}$ No. of cores} & \multicolumn{1}{c|}{2 x 40}              & 2 x 56          & 1 x 112                   \\
\textbf{CPU Clock Rate}                      & \multicolumn{1}{c|}{2.4 GHz}             & 2GHz            &                           \\
\textbf{Memory} & \multicolumn{1}{c|}{256 GB DDR5} & 512 GB DDR5 & 512 GB DDR4 \\
\textbf{Network}                              & \multicolumn{1}{c|}{4x NDR200}           & 1x NDR200  (shared by 2 nodes)     & 2xNDR200                 \\
\textbf{Node Bandwidth}     & \multicolumn{1}{c|}{800Gb/s}                  & 100Gb/s            &   400Gb/s                 \\
\textbf{Compiler / version}                          & \multicolumn{1}{c|}{mpicxx / nvc++ 23.11-0}               & mpicxx / nvc++ 23.11-0               & mpicxx / nvc++ 24.5-1                     \\
\textbf{No. of GPUs per node}                     & \multicolumn{1}{c|}{4}                   & –               & 4                         \\
\textbf{GPU Model}                           & \multicolumn{1}{c|}{NVIDIA Hopper H100 SXM5}    & –               & NVIDIA custom Ampere A100 \\
\textbf{GPU Memory}                          & \multicolumn{1}{c|}{HBM3 64 GB}               & –               & HBM2e 64 GB                     \\  

\textbf{HBM Bandwidth}                           & \multicolumn{1}{c|}{3.35 TB/s}       & –               &  2.0 TB/s \\
\textbf{FP64 (Double Precision)}                           & \multicolumn{1}{c|}{34 TFLOPS}       & –               &  9.7 TFLOPS \\ \midrule
\bottomrule
%\textbf{Driver version}                          & \multicolumn{1}{c|}{...}               & ...               & ...   \\\midrule\bottomrule

\end{tabular}
\caption{Technical specification of the Marenostrum Accelerated and General Purpose Partition and Leonardo Booster.}
\label{tab:clusters}
\end{table*}

\end{document}